\documentclass[aps,pra,superscriptaddress,twocolumn,nofootinbib
]{revtex4-2}

\usepackage{graphicx}
\usepackage{amsmath}
\usepackage{amssymb}
\usepackage{amsfonts}
\usepackage{amsthm}
\usepackage{hyperref}
\usepackage{color}
\usepackage{xcolor}
\usepackage{soul}
\usepackage{bm}
\usepackage{dsfont}
\usepackage{bbm}
\usepackage{braket}
\usepackage{enumitem}   
\usepackage{wrapfig}
\usepackage{cleveref}

\usepackage{physics}

\usepackage[draft]{changes}
\usepackage{comment}
\newtheorem{assumption}{Machine Requirement}
\newtheorem*{result}{Result}

\renewcommand{\Tr}[1]{\text{Tr}\left[#1\right]} 

\newcommand{\J}{\mathbb{J}}

\newcommand{\F}{\mathcal{F}}
\newcommand{\D}{{\rm d}}

\newcommand{\id}{\mathbb{I}}
\newcommand{\llangle}{\langle\!\langle}
\newcommand{\rrangle}{\rangle\!\rangle}
\newcommand{\h}{\kappa}
\renewcommand{\epsilon}{\varepsilon}

\begin{document}

\author{Paolo Abiuso}
\affiliation{Institute for Quantum Optics and Quantum Information - IQOQI Vienna,
Austrian Academy of Sciences, Boltzmanngasse 3, A-1090 Vienna, Austria}

\author{Alberto Rolandi}
\affiliation{Atominstitut, TU Wien, 1020 Vienna, Austria}

\author{John Calsamiglia}
\affiliation{F\'isica Te\`orica: Informaci\'o i Fen\`omens Qu\`antics, Department de F\'isica, Universitat Aut\`onoma de Barcelona, 08193 Bellaterra (Barcelona), Spain}

\author{Pavel Sekatski}
\affiliation{D\'{e}partement de Physique Appliqu\'{e}e,  Universit\'{e} de Gen\`{e}ve,  1211 Gen\`{e}ve,  Switzerland}

\author{Martí Perarnau-Llobet}
\affiliation{F\'isica Te\`orica: Informaci\'o i Fen\`omens Qu\`antics, Department de F\'isica, Universitat Aut\`onoma de Barcelona, 08193 Bellaterra (Barcelona), Spain}
\affiliation{D\'{e}partement de Physique Appliqu\'{e}e,  Universit\'{e} de Gen\`{e}ve,  1211 Gen\`{e}ve,  Switzerland}

\title{An information-theoretic proof of the Planckian bound for thermalization
}

\begin{abstract}
   We demonstrate that quantum mechanics entails a fundamental lower bound on the thermalization time $\tau$ of any system. At finite temperature, we show that $\tau$ is bounded by half the Planckian dissipation time, $\tau \geq \tau_{\rm Pl}/2$ with  $\tau_{\rm Pl} = \hbar/(k_{\rm B} T)$. 
   In the low-temperature regime, our bound takes the form  $\tau \geq \hbar / \Delta$ with $\Delta$ the spectral gap, in close connection with  the quantum adiabatic theorem. These bounds, rooted in Hamiltonian estimation, 
   hold for arbitrary  quantum  processes that output states close to the corresponding thermal ensemble for a nontrivial class of Hamiltonians. 
\end{abstract}

\maketitle


\section{Introduction}
Understanding equilibration and thermalization from the underlying reversible quantum dynamics is one of the most fundamental, long-lasting open questions in physics~\cite{Polkovnikov2011Nonequilibrium,Gogolin2016,Linden2009Quantum,Garcia2017Equilibration}. 
One profound aspect of thermalization in quantum many-body systems is the
emergence of the Planckian dissipation time $\tau_{\rm Pl}$~\cite{hartnoll2022colloquium,nussinov2022exact}, as given by  
\begin{equation}
    \tau_{\rm Pl} = \frac{\hbar}{k_{\rm B} T}. 
    \label{eq:planckiantime}
\end{equation}
This universal
timescale grows inversely with temperature $T$ and is determined by two fundamental constants in nature, namely Planck’s constant $\hbar$ and Boltzmann’s  constant~$k_B$. 
Analogously to the ``Planck time" in quantum
gravity, it is conjectured to represent the shortest timescale for thermalization. That is, the minimum time required to reach a Gibbs state starting from an arbitrary non-equilibrium distribution~\cite{Wilming2018,hartnoll2022colloquium,Goldstein_2015,Reimann2016,lucas2019operator,Vikram2024Exact}. 

Originally, the concept of a ``Planck scale of dissipation", as given by \eqref{eq:planckiantime}, was coined by Zaanen to describe the universal behavior of the conductance of superconductors above their critical temperature~\cite{zaanen2004why,Zaanen2019}. This behavior was comprehensively empirically corroborated in~\cite{bruin2013similarity}, which ignited the quest for an underlying microscopic explanation, tentatively put forward for metals in~\cite{hartnoll2015theory,mousatov2020planckian}, and more recently in~\cite{delacretaz2025bound}, where a bound is provided for the emergence of hydrodynamic 
behavior in many-body systems. 
In parallel, the concept of Planckian time has been linked to the rate of growth of chaos in many-body systems and  field theories~\cite{Maldacena2016,Tsuji2018Bound,Murthy2019Bounds,Pappalardi2022,Delacretaz2022}. 

Despite the omnipresence of the Planckian time as a fundamental limit on the speed of dissipative
or chaotic processes~\cite{hartnoll2022colloquium,lucas2019operator}, it is not difficult to come up with
apparent violations of this conjecture. 
For instance, in collisional models, thermalization is described by letting the system interact with a bath made up of identical copies of the system in thermal equilibrium~\cite{Scarani2002}. By letting these collisions happen arbitrarily fast, one realises there is no fundamental restriction on the time required to relax to the thermal state of a \emph{previously known} Hamiltonian. 
The best one can hope for are bounds assuming a limited interaction strength~\cite{Shiraishi2021Speed}.

Motivated by this tension, in this work we put forward a new approach to rigorously prove the appearance of the Planckian bound for thermalization based on quantum information geometry and
metrology. Unlike previous approaches, our assumptions on the structure of the quantum system and the allowed dynamics are minimal and model-independent. 

Our starting point is to introduce the concept of a   \emph{thermalization machine} $M$, namely a device thermalizes 
a quantum system $S$. We make the following basic assumptions on the action of the machine:
\begin{itemize}
     \item[R1] {\bf Validity of Quantum Mechanics}. The microscopic evolution of the system and the machine (including the environment for open systems) is well described by the Schrödinger equation. 
    \item[R2] {\bf Thermalization}. The machine~$M$ should output states that are close to the thermal ensemble, independently of the specific structure of~$S$. 
\end{itemize}
The second requirement 
is crucial as it makes a clear distinction between state preparation and (chaotic) thermalization processes. 
In this framework, standard thermal baths satisfying detailed balance~\cite{breuer2002theory} as well as complex many body system satisfying the ETH hypothesis~\cite{DAlessio2016,Deutsch_2018} would be considered thermalizing machines. 

Under these two minimal assumptions, we provide a universal bound~\eqref{eq:chi_def} on the time $\tau$ required to reach thermalization at inverse temperature $\beta=1/k_{\rm B}T$. When considering thermalization in a system with a spectral gap $\Delta$---defined as the energy difference between the ground state and the first excited state---our bound can be summarized as follows
\begin{equation}
    \tau \geq
    \begin{cases}
        \tau_{\rm Pl}/2, & \beta \Delta \lesssim 1 \\
        \hbar/\Delta, &  \beta \Delta \gg 1. 
    \end{cases}
    \label{eq:ourboundintro}
\end{equation}
This is in good agreement with previous results in the literature, as the Planckian dissipation behavior is known to break down close to zero temperature~\cite{hartnoll2022colloquium}. 



\section{Setting the stage: An information-theoretic notion of  thermalization}

Inspired by quantum information theory, our framework is adversarial: it assumes the existence of an entity (the machine $M$) that has full control to realize a desired task (thermalization), in the shortest possible time~$\tau$, under minimal physical constraints.
In order to minimize~$\tau$, we allow \emph{any} control on $M$ and its interaction with $S$, though they  should---to some extent---be independent of the specifics of~$S$. This is formalized through two main requirements on the thermalizing machine.


\begin{assumption}[Validity of Quantum Mechanics]
\label{ass:q_mech}
The evolution of $S$ and $M$ is  described by quantum mechanics 
\begin{equation}
\begin{split}
     \,  \dot \rho_{SM}(t)&= -\frac{i}{\hbar}[ H_S+V_{SM}(t)+H_M(t), \rho_{SM}(t)] \\  
    \rho_S(t,H_S) &:= \tr_M \rho_{SM}(t) 
  \end{split}
   \label{eq:stateevolution}
\end{equation}
where $H_S$ is the system Hamiltonian, while $H_M(t)$ ($V_{SM}(t)$)  the machine (interaction) Hamiltonian which specify the machine. 
\end{assumption} 

\begin{figure}
    \centering \includegraphics[width=0.9\linewidth]{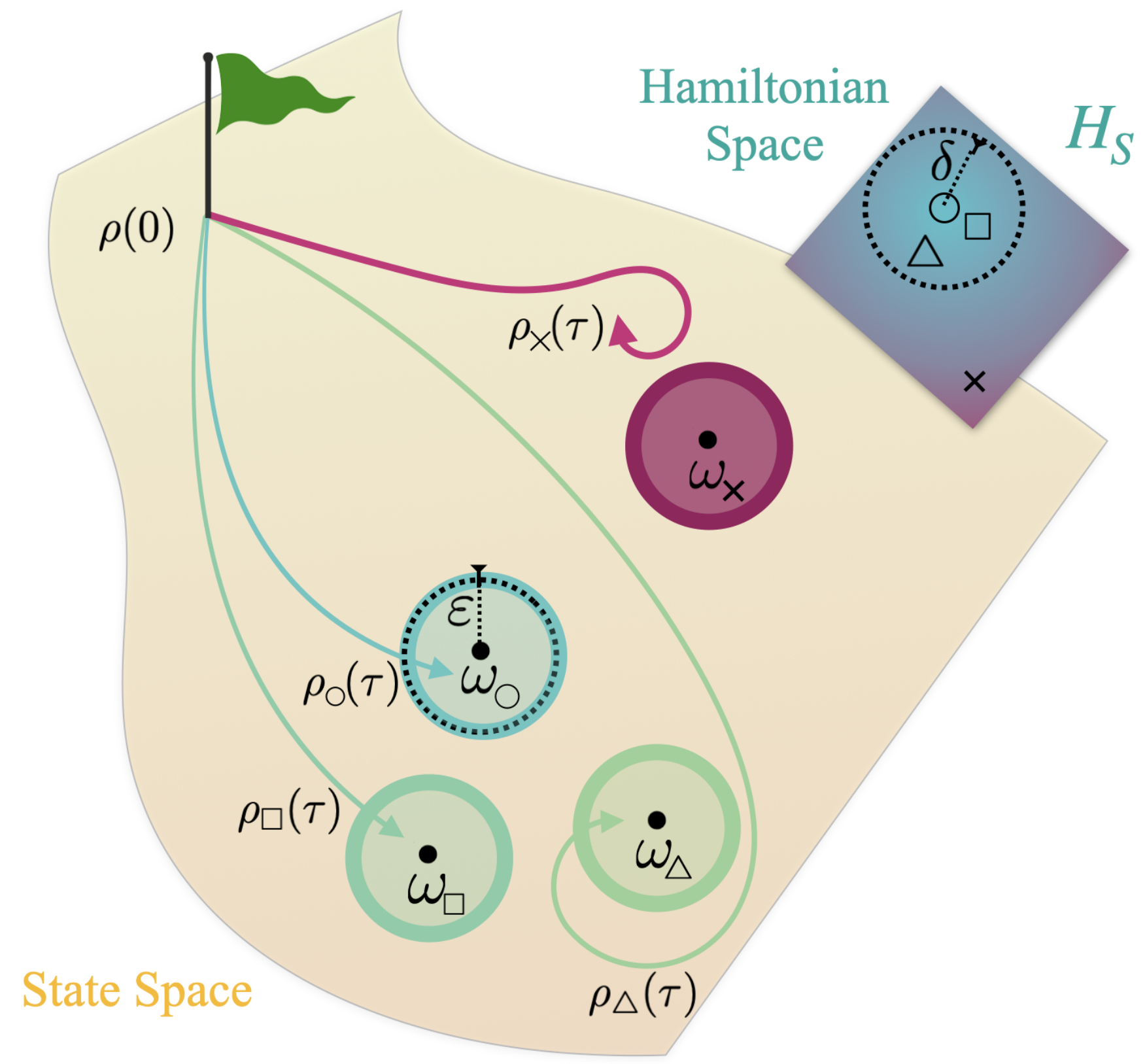}
    \caption{
    {\bf Representation of thermalization process for a given machine $M$.} Depending on the local Hamiltonian $H_S$, the dynamical evolution induced by $M$ is required to bring the system $S$ close to the corresponding thermal state after a time $\tau$. This should happen  up to an error $\varepsilon$ and for a nontrivial subset of Hamiltonians (here a ball of size $\delta$).}
    \label{fig:traject}
\end{figure}

Note that for open quantum systems we include all of the environment into the machine, in other words we consider any time-dependent unitary dilation of the dynamics. We also emphasize that there is no restriction on the size of $M$ and the strengths of $H_M$ and $V_{SM}$, i.e. the machine is free to do anything allowed by quantum mechanics. 
This machine should be able to thermalize the system $S$ {\it for at least some choices of $H_S$}, which is illustrated in Fig.~\ref{fig:traject} and formalized as follows:


\begin{assumption}[
Thermalization]
\label{ass:local_term}
For some initial state $\rho_{SM}(0)$ 
and after time~$\tau$, the machine outputs states~$\rho_S(\tau,H_S)$ that are close to the corresponding thermal ensemble at temperature~$T=(k_{\rm B}\beta)^{-1}$, for all choices of $H_S$ from some set. Specifically for a ball $\mathcal{B}_\delta$ around $\bar H_S$ we require 
\begin{align}
       &(i)\quad  D\big(\rho_S(\tau,H_S),\omega(\beta,H_S)\big)
        \leq \varepsilon
        \label{eq: approx_thermalization}
        \\
        \label{eq:smooth_thermalization}
         &(ii)\quad  \forall \, H_S\in\mathcal{B}_\delta \text{ with } \mathcal{B}_\delta:=\{H\big| \|H-\bar{H}_S\|\leq \delta\}\;,
\end{align}
where $\omega(\beta,H_S):= e^{-\beta H_S}/\Tr{e^{-\beta H_S}}$ are Gibbs states, $D(\rho,\sigma):= \arccos(\sqrt{F(\rho,\sigma)})$ is chosen to be the Bures angle ($F$ is the quantum fidelity), and $\|A\|:=\lambda_{\rm max}(A)-\lambda_{\rm min}(A)$ is the spectral semi-norm, see Sup.Mat.~\ref{app:math_preamble} for details. The parameters $\delta$ and $\epsilon$ control the range  and accuracy of the thermalization machine, respectively.
\end{assumption}

We claim that Requirement~\ref{ass:q_mech} and ~\ref{ass:local_term} are minimal criteria that should be satisfied to prove the existence of a minimal thermalization timescale:
1) Abandoning Requirement~\ref{ass:q_mech} 
and assuming the validity of classical mechanics at all scales ($\hbar\rightarrow 0$), would take the Planckian time to zero.  
2) Requirement \ref{ass:local_term} is essential to distinguish a thermalization machine from a single-state preparation one. Indeed, a collisional ``swap machine'' can prepare a fixed $\omega(\beta,\Bar{H}_S)$ instantaneously, as  discussed above. 

Finally, it is worth mentioning that both conditions $(i)$ and $(ii)$ are chosen for simplicity and can be relaxed. In particular, for the derivation of our results it is in general sufficient to require an even weaker notion of thermalization, in which $(i)$ is satisfied for  at least two different $H_S$. Further, it is also sufficient to require $(i)$ to be valid for some weighted-average of $\rho(t,H_S)$ with $0\leq t \leq \tau$ (this includes the possibility of a finite time-resolution of the devices), or only after dephasing e.g. in the local energy eigenbasis of $S$.
For a detailed discussion see the Sup.Mat.~\ref{app:formal_derivation}.


\section{The Planckian bound on thermalization}

We demonstrate the following lower bound on the thermalization time.

\begin{result}
\label{res:main}
For any thermalization machine satisfying the requirements~\ref{ass:q_mech} and~\ref{ass:local_term} the thermalization time must satisfy 
\begin{align}
   & \tau \geq   \tau_{\rm Pl} \cdot \chi( \bar{H}_S,\delta,\varepsilon)\;, \quad \quad \text{with \ }   
\label{eq:chi_def}\\
  &\chi( \bar{H}_S,\delta,\varepsilon):=\hspace{-1mm}\max_{ H^{(1,2)}_S \in \mathcal{B}_\delta}
 \hspace{-1mm} \left[\frac{2D\big(\omega(\beta,H^{(1)}_S),\omega(\beta,H^{(2)}_S)\big)-4\varepsilon}{\beta  \hspace{1mm} \|H^{(1)}_S-H^{(2)}_S\| } \right]
 \nonumber
\end{align}   
\end{result}
This result can be understood as a universal bound on the thermalization time. Indeed, as we show below, the adimensional factor $\chi( \bar{H}_S,\delta,\varepsilon)$ in Eq.~\eqref{eq:chi_def} is finite and remains bounded from zero in all regimes where thermalization (requirement~\ref{ass:local_term}) remains sufficiently different from single-state preparation. One also notes the upper bound $\chi( \bar{H},\delta,\varepsilon)\leq \frac{1}{2} - \frac{4\epsilon}{\beta \delta}$ highlighting that accuracy and range parameters must satisfy $\epsilon\leq \frac{\beta \delta }{8}$ in order to yield a nontrivial bound (in other words, the tolerated error should be sufficiently small to distinguish the different thermal states required).

{\it Proof sketch:} 
The core idea of the proof is that the distinguishability between the outputs of the thermalization machine generated by different system Hamiltonians $H_S$ is fundamentally constrained by the sensitivity of the global unitary evolution to changes in $H_S$ \eqref{eq:stateevolution}.
This can be concretized in
information-geometrical arguments (see Sup.Mat.~\ref{app:formal_derivation} for full details):  first, when requirement~\ref{ass:local_term} holds for two Hamiltonians $H_S^{(1)}$ and $H_S^{(2)}$, the triangle inequality for the Bures angle implies
$D\left(\rho_S(\tau,H_S^{(1)}),\rho_S(\tau,H_S^{(2)})\right) \geq D\left(\omega(\beta,H_S^{(1)}),\omega(\beta,H_S^{(2)})\right)-2\epsilon$.
In turn, requirement~\ref{ass:q_mech} sets a limit on variations of system $S$ under variations of the local $H_S$ -- after time $\tau$ the possible states of the system must satisfy  
$D\left(\rho_S(\tau,H_S^{(1)}),\rho_S(\tau,H_S^{(2)})\right) \leq \frac{\tau}{2\hbar} \|H_S^{(1)}- H_S^{(2)}\|$.
The result~\eqref{eq:chi_def} is then obtained by optimizing the choice of the Hamiltonians. $\square$

In what follows, we will characterize $\chi(\bar{H}_S,\delta,\epsilon)$  in different physically relevant limits to obtain universal bounds on thermalization, and contrast such bounds with explicit dynamics/machines in sec.~\ref{sec:models} and \ref{subsec:resonant_model}.

\subsection{Locally-exact thermalization and Quantum Fisher Information}
\label{sec:fisher_limit}

Let us now consider the case of \emph{locally-exact thermalization}, letting  $\frac{\epsilon}{\beta} \ll \delta \to 0$ in the Requirement~\ref{ass:local_term}. In this limit the machine must prepare the exact Gibbs state $\rho_S(\tau,H_S)\equiv \omega(\beta,H_S)$, for all \emph{perturbations} of the Hamiltonian $H_S(\delta,\kappa) = \bar H_S + \delta \kappa$ with $\|\kappa\|\leq 1$ and infinitesimal $\delta$. Then, the bound~\eqref{eq:chi_def} becomes
\begin{align}
   \hspace{-2mm}  \tilde{\chi}(\bar{H}_S)
    \label{eq:boundQFI}:=\lim_{\frac{\varepsilon}{\beta}\ll \delta\rightarrow0}\chi( \bar{H}_S,\delta,\varepsilon) = \max_{\h=\h^\dagger} \frac{\sqrt{\mathcal{F}_\h^{\rm th}(\beta, \bar{H}_S)}}{\beta\|\h\|},
\end{align}
where $\mathcal{F}_\h^{\rm th}(\beta, \bar{H}_S) \equiv  \mathcal{F}\big( \omega(\beta, H_S(0,\kappa))\big)$  is the  quantum Fisher information (QFI) of a thermal state~\cite{paris2009quantum,abiuso2025fundamental}. Here, we used the fact that the QFI of a parametric state $\rho_\delta$ is related to its susceptiblity with repect ot the Bures angle via $\F(\rho_0) = 4\left( \lim_{\delta\to 0}\frac{D(\rho_0,\rho_{\delta})}{\delta} \right)^2$  (cf. Sup.Mat.~\ref{app:math_preamble}).

This result admits a natural interpretation in terms of quantum metrology.  For locally-exact thermalization, $\mathcal{F}_\h^{\rm th}(\beta, \bar{H}_S)$ must coincide with the ``dynamical" QFI $\mathcal{F}_\h^{\rm dyn}(\tau, \bar H_S) \equiv  \mathcal{F}\big( \rho_S(\tau, H_{S}(0,\kappa)) \big)$, which is bounded by the generalized Heisenberg limit $\mathcal{F}_\h^{\rm dyn}(\tau, \bar{H}_S)\leq \|\h\|^2 \tau^2 /\hbar^2$~\cite{Boixo2007}. By maximizing over the Hamiltonian perturbations $\kappa$, we then immediately recover Eq.~\eqref{eq:boundQFI}. In words, this bound follows from the observation that the thermalization machine can not violate the Heisenberg limit.

The maximization in \eqref{eq:boundQFI} is carried out in the Sup. Mat.~\ref{sec:fisherbounds}. There, we show that $\tilde{\chi}\geq \sqrt{p(1-p)}$ for all possible bipartitions of the population of $\omega(\beta,\bar{H}_S)$ in two sets
with probabilities~$\{p,1-p\}$.
When  $\omega(\beta,\bar{H}_S)$ is sufficiently mixed so that $p \approx 1/2$ can be chosen, we find $\tilde{\chi} \approx 1/2$, recovering the first line of Eq.~\eqref{eq:ourboundintro}. In particular,  $\tilde{\chi}\geq \sqrt{2}/3 \sim0.47$ whenever the ground state probability $p_0$ is below $2/3$.
In the opposite regime $p \to 1$, in which  $\omega(\beta,\bar{H}_S)$ approaches the ground state, we derive a different bound: $\tilde{\chi} \geq (2p_0 - 1)/(\beta \Delta)$ for $\beta \Delta \gg 1$, where $\Delta$ is the spectral gap. This leads to the second line of Eq.~\eqref{eq:ourboundintro}.


\begin{figure}
    \centering
    \includegraphics[width=\linewidth]{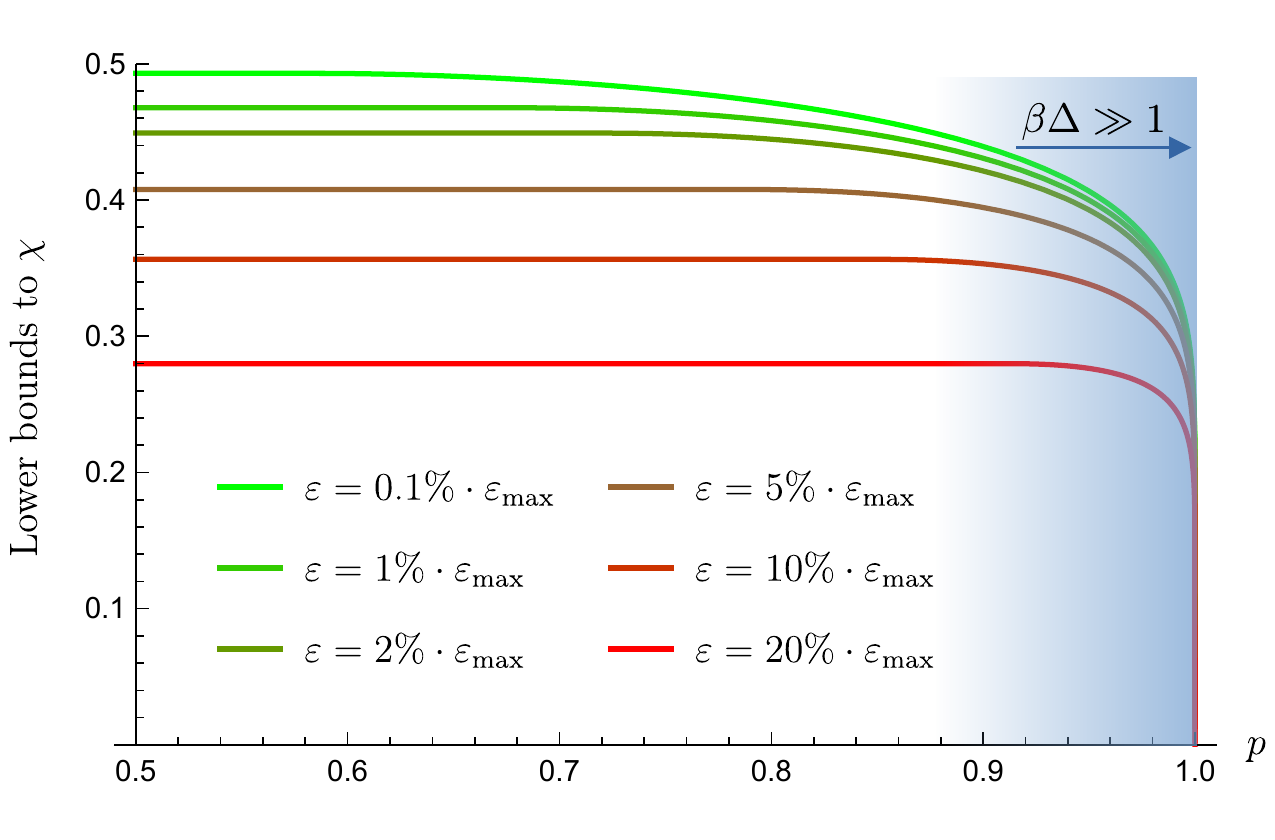}
    \vspace{-15pt}
    \caption{
    {\bf Bounds for approximate thermalization.}  For different finite values of the tolerated error $\varepsilon$, lower bounds to $\chi(\bar{H}_S,\delta,\varepsilon)$ are shown as a function of any bipartition $\{p,1-p\}$ of the thermal state populations at $\bar{H}_S$. Specifically, \eqref{eq:chi_def} is partially optimized over a simple set of Hamiltonians $H_S^{(i)}$, namely those commuting with $\bar{H}_S$ and such that $H_S^{(i)}-\bar{H}_S$ has only two distinct eigenvalues. In units of Bures angle the maximum tolerated error corresponds to $\varepsilon_{\rm max}\equiv\pi/4$. For $\varepsilon\lesssim 5\%\cdot \epsilon_{\rm max} $, $\chi$ is in general at least larger than $\sim0.4$. In the limit $p\rightarrow 1$ (i.e. close to ground state=) we find $\chi$ to tend to zero strictly slower than $1/(\beta\Delta)$ (details in Sup.Mat.).}
    \vspace{-15pt}
    \label{fig:main_lowerbounds}
\end{figure}

\subsection{Approximate thermalization}
The limit of locally-exact thermalization  yields simple bounds and an intuitive understanding in terms of Fisher information. The more general inequality~\eqref{eq:chi_def} follows from a similar geometrical argument using finite variations of $H_S$ while admitting the possibility of a finite error $\varepsilon >0$ in reaching the exact thermal state. Such possibility is crucial to ensure the continuity and, more importantly, the wide validity of our main results: thermalization in nature is not, in general, exact.

On the formal level, moving away from $\epsilon=0$ makes the function $\chi(\bar{H}_S, \delta, \varepsilon)$~\eqref{eq:chi_def} more challenging to compute. Remarkably, we can compute again different lower bounds that hold for any $\bar{H}_S$ in any Hilbert space dimension and only depend on the possible bipartite coarse-graining of the thermal states population.
These bounds are shown in Figure~\ref{fig:main_lowerbounds} 
and we observe once again how they yield $\chi\gtrsim 0.5$ for states that are sufficiently mixed (i.e. when $\omega(\beta,\bar{H})$ is not concentrated in the ground state only), and sufficiently small error. As the error $\varepsilon$ increases, the machine might in principle become faster, however notice that at $\sim 5\%$ error one still has $\chi\gtrsim 0.4$, and  $\chi\gtrsim 0.3$ at around $20\%$ error. 
Remarkably, all lower bounds in Figure~\ref{fig:main_lowerbounds} can be obtained by considering a simple subset of Hamiltonians that are diagonal in the basis of $\bar{H}_S$. That is, they hold even when $M$ is only required to thermalize classical (commuting) Hamiltonians. 



\subsection{Approaching the Planckian limit} 
\label{sec:models}


After deriving and analyzing the fundamental inequality~\eqref{eq:chi_def}, we now address its tightness, that is, are there specifically-engineered thermalization machines that can approach the Planckian time? 
Since our bound is a consequence of the machine's inability to thermalize two different Hamiltonians in a given time, a preliminary attempt
is 
offered by Hamiltonian discrimination, where it is known that two Hamiltonians $H_S^{(1)}$ and $H_S^{(2)}$ can be perfectly distinguished after time $\tau$ if an only if $\tau\geq \hbar \pi/\|H_S^{(1)}-H_S^{(2)}\|$~\cite{aharonov2002measuring}. Clearly, this time is also sufficient for thermalization, since preparing the Gibbs state of a known Hamiltonian is trivial in our framework. This ``discrimination" time however is longer than our lower bound~\eqref{eq:chi_def}, and is even diverging for nearby Hamiltonians. Inspired by this discrimination task, we now describe a machine which saturates the Planckian bound for any given pair of Hamiltonians.

First, note that after time $\tau$ the machine used for Hamiltonian discrimination~\cite{aharonov2002measuring} 
prepares the system in two pure states $\ket{\psi_i(\tau)}_S$, corresponding to the Hamiltonians $H_S^{(i)}$, such that  $|\braket{\psi_1(\tau)}{\psi_2(\tau)}|^2 = \cos^2\left(\frac{\tau}{2 \hbar } \|H_S^{(1)}-H_S^{(2)}\|\right)$ (see Sup.Mat.~\ref{app: opt two} for details).
With an isometry $V$ the machine can then map these two states to any pair of pure states $\ket{\Psi_i}_{SM}= V\ket{\psi_i(\tau)}_S$
that are purifications of $\omega(\beta, H^{(i)})$. By Uhlmann's theorem, we can choose these states so that they satisfy $|\braket{\Psi_1}{\Psi_2}|^2 = F(\omega(\beta, H^{(1)}),\omega(\beta, H^{(2)}))$. Hence, our machine can thermalize the two Hamiltonian after time 
\begin{equation}
    \tau = \frac{2 \, \hbar\, D(\omega(\beta, H^{(1)}),\omega(\beta, H^{(2)}))}{\|H_S^{(1)}-H_S^{(2)}\|},
\end{equation}
saturating the bound~\eqref{eq:chi_def} for $\epsilon=0$.


\subsection{A case study: Resonant-level model of thermalization}
\label{subsec:resonant_model}
\begin{figure}
    \centering
    \includegraphics[width=\linewidth]{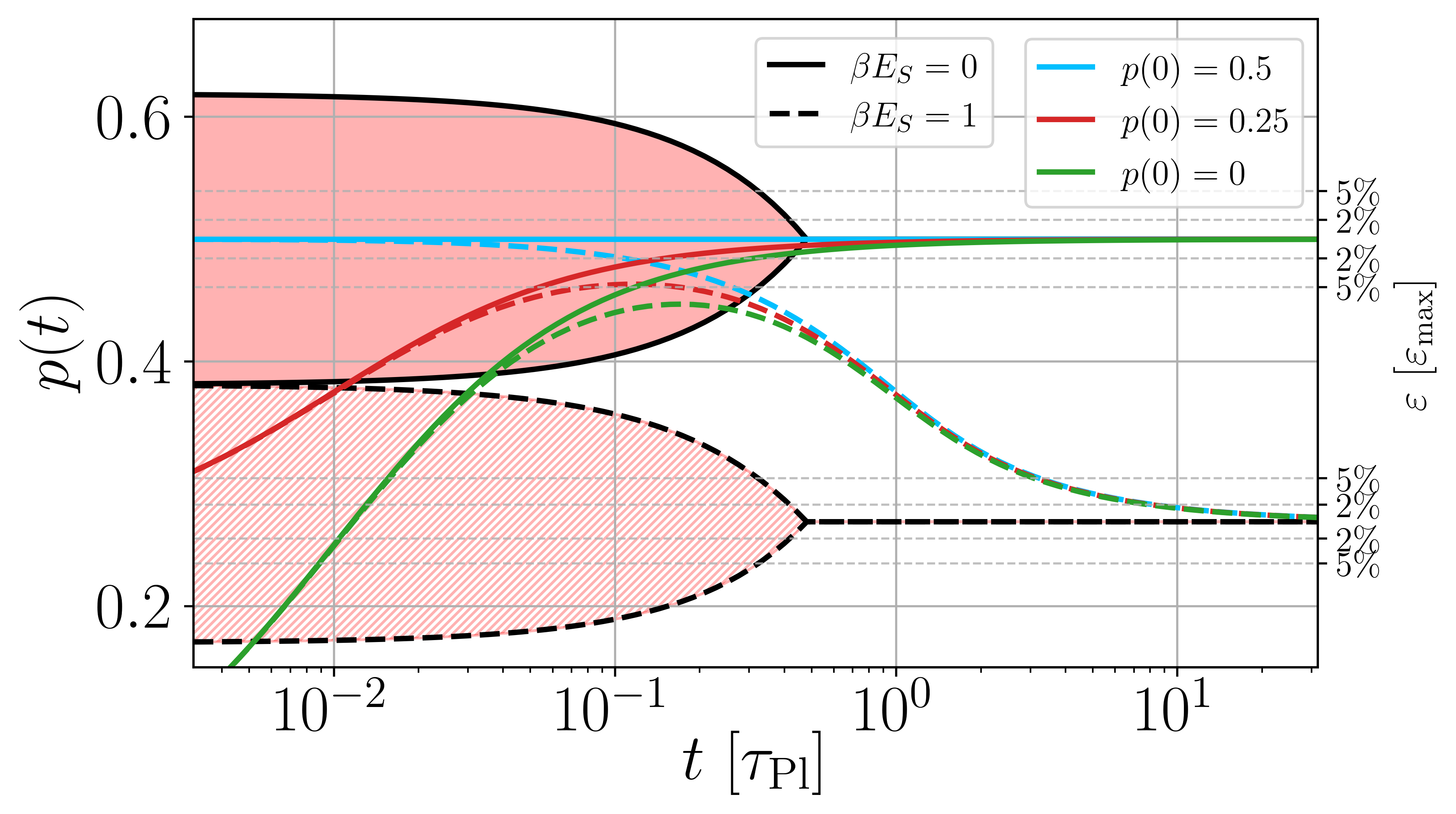}
    \vspace{-20pt}
    \caption{{\bf Exact evolution of a fermionic model vs. Planckian thermalization bound}. For each value of $\varepsilon$ our bound~\eqref{eq:chi_def} sets a minimum time $\tau$ under which a machine is unable to thermalize different Hamiltonians simultaneously. For the same initial condition, if a given machine thermalizes quickly one Hamiltonian $H_S^{(1)}$ (plain red area) then it cannot thermalize a second Hamiltonian $H_S^{(2)}$ (dashed red area) with arbitrary precision in the same amount of time. This must hold for any conceivable machine $M$  that is able to thermalize $S$. Here, we illustrate it for the fermionic machine described in the text ($p(t):=\Tr{\rho_S(t) \hat{a}^\dagger \hat{a}}$).
    As it can be seen from the plot, the evolution is such that one of the two Hamiltonians is thermalized faster than the Planckian time (plain line in plain red area or dashed line in dashed red area) then the other one is not. Only after a time of the order of the Planckian time has passed we can see that both Hamiltonians can thermalize.
    }
    \vspace{-20pt}
    \label{fig:rlm_fig}
\end{figure}
The above example shows how a fine-tuned machine could, in principle, approach the Planckian limit of thermalization. We now discuss   a physical model where  $S$ is a single fermion described by (commuting) Hamiltonians $H_S = E_S\hat{a}^\dagger \hat{a}$, 
and thermalization can be characterized through its occupation probability  $p(t):=\Tr{\rho_S(t) \hat{a}^\dagger \hat{a}}$. 
Starting from the same initial state, the bound from Eq.~\eqref{eq:chi_def} forbids any machine to concurrently thermalize multiple Hamiltonians in less than $t=\tau_{\rm Pl}\cdot\chi$. That is,
when the evolution with a Hamiltonian $H_S^{(1)}$ is too close to the thermal state $\omega(\beta,H_S^{(1)})$, then the evolution with another Hamiltonian $H_S^{(2)}$ has to be far from $\omega(\beta,H_S^{(2)})$, and vice-versa. In Fig.~\ref{fig:rlm_fig} we visualize this with the red regions by picking two Hamiltonians $E_S^{(1)} = 0$ and $E_S^{(2)}=k_BT$. 
At $t=0$ the size of these forbidden areas (in red) is maximal, and as time increases to the order of the Planckian time the size decreases to zero, in the figure the bound becomes trivial at $t\approx\tau_{\mathrm{Pl}}/2$.

\begin{table*}[]
    \centering
    \begin{tabular}{c|c|c|c}
         & Gibbs state sampling & Thermalization machine & Hamiltonian learning  \\
         \hline
        \emph{Input} & Exact knowledge of $H$ & Oracle access to $H$ & Oracle access to $H$  \\
        \emph{Desiderata} & Thermal statistics & Thermal statistics & $H$ discrimination \\
        \emph{Minimum time} (for 2 values of $H$) & 0 & $\tau_{\rm Pl}/2$ [this work] & $\tau_{\rm AMP}$~\cite{aharonov2002measuring}
    \end{tabular}
    \caption{{\bf Information protocols based on Hamiltonian knowledge.} Comparison among Gibbs state sampling protocols, thermalization machines, and Hamiltonian learning: GSS requires reproducing the thermal statistics induced by some Hamiltonian $H$  given its classical description. In the case of HL the goal is more demanding -- learning $H$ -- and the premise is weaker -- some kind of oracle-access to $H$. In this context, our thermalization machines share the \emph{desiderata} with GSS, and the \emph{input} with HL.
    In the table, the information provided to the machine decreases from left to right, while the output information required increases. As a consequence, these tasks are increasingly harder, and the fundamental minimum time needed by an all-powerful entity to accomplish these tasks is ordered accordingly; when only 2 possible values of $H$ are given, and the entity is only limited to obey quantum mechanics, GSS can in principle be arbitrarily fast -- via a control-swap machine -- whereas thermalization and HL have to comply with the Planckian bound~\eqref{eq:chi_def} and the Hamiltonian discrimination bound~\cite{aharonov2002measuring}.
    }
    \label{tab:comparison}
\end{table*}

As a testing playground, we choose a machine $M$ consisting of a fermionic bath with  Hamiltonian $H_M = \sum_k \Omega_k \hat{b}^\dagger_k \hat{b}_k$ that can interact with $S$ via tunneling interactions $V_{SM}(t) = g(t)\sum_{k=1}^n \lambda_k \hat a^\dagger \hat b_k + \lambda_k^* \hat b_k^\dagger \hat a$, the so-called resonant level model~\cite{schaller2014open,Rolandi2023finitetimelandauer}.  The time-dependent function $g(t)$ characterizes the machine's operating schedule.
For initially uncorrelated $S$ and $M$ 
we can solve the exact dynamics in the thermodynamic limit (cf. Supp. Mat.~\ref{app:rlm}) for any choice of local energy gap $E_S$ and coupling $g(t)$. 
In particular we analyze in details schedules of the form $g(t) = \sqrt{a/(t+b)}$ for $0<t<\tau$, $a,b>0$. We observe in Fig.~\ref{fig:rlm_fig} that these trajectories approach the Planckian limit, while satisfying our constraints.

\section{Discussion}
We presented a fundamental time-bound~\eqref{eq:chi_def} that is to be satisfied by any thermalization process in quantum mechanics.
In order to formalize our results in the most general framework, we introduced the concept of \emph{thermalization machine} $M$, which only has to comply with quantum mechanics and output states close to thermal equilibrium. 
This concept connects two tasks that have been intensively studied in recent years, 
namely Gibbs state sampling (GSS)~\cite{poulin2009sampling,temme2011quantum,chen2023efficient,gilyen2024quantumgeneralizations,Chen2024Boosting,Rouze2024Optimal} and Hamiltonian learning (HL)~\cite{bairey2019learning,huang2023learning,dutkiewicz2024advantage}, see Table~\ref{tab:comparison}. 

Our results can be summarized as~\eqref{eq:ourboundintro}: 
if $M$ is required to thermalize \emph{all} Hamiltonians $H_S$, or at least those leading to thermal states that are sufficiently mixed ($\beta\Delta\lesssim 1$), $\tau_{\rm Pl}/2=\hbar/(2k_{\rm B}T)$ sets the universal minimum operational time of $M$.
When the 
the machine is only required to be a \emph{ground state cooling machine}, its minimum operational time is bounded by (less stringent)~$\hbar/\Delta$, in correspondence to optimal protocols for adiabatic state preparation \cite{Albash2018}.
The core idea behind these results consists in comparing the distinguishability of thermal states with the global variation of unitary dynamics induced by their Hamiltonians. 

While being fully model-independent, our approach can be tailored to more specific physical scenarios once additional details are known about $S$ or $M$. 
This calls for further questions: in multipartite systems, what is the role of entanglement in the Planckian bound? What happens in the large-size limit? How close can real-life thermal machines approach the Planckian bound? While a comprehensive answer to these questions is beyond the scope of this work and motivation to further investigation, let us provide some insight here.

Taking the limit of locally-exact thermalization considered in~\ref{sec:fisher_limit}, let us suppose that the user of the machine is not able to do full tomography of the output. Or rather, that they can only measure some observable $A=\sum_a a\Pi_a$. Then, the accessible information of the user is limited to 
$p_a=\Tr{\Pi_a \omega(\beta,\bar{H})}$.
By definition, the corresponding accessible Fisher information is upperbounded by the thermal QFI: $\sum_a \frac{({\partial_\theta p_a})^2}{p_a}\leq \F^{\rm th}_{\partial_\theta H}$. 
Then, following the same steps delineated in Sec.~\ref{sec:fisher_limit} one can bound such observable Fisher information with the QFI of the unitarily-evolving $S+M$.
In case the system $S$ is $N$-partite one finds 
$
    \tau \gtrsim \beta\hbar \sqrt{N^{\alpha-\phi}},
$
where $N^\alpha$ represents the scaling of the 
thermal QFI of $S$ and $N^{\phi}$ of the global dynamical QFI of $S+M$.
Standard uncorrelated thermal systems satisfy $\alpha=1$, while it has been proven in~\cite{abiuso2025fundamental} that local classical observables can achieve up to $\alpha=2$ on strongly correlated thermal systems.
Moreover, $\phi=2$ needs the machine dynamics to generate consistent $N$-partite entanglement in $\rho_{SM}$, 
whereas $\phi=1$ when a separable partition can  be found at all times (see Sup.Mat.~\ref{app:dynQFI}).

\acknowledgements

P.A. acknowledges fundings from the Austrian Science Fund (FWF) projects I-6004 and ESP2889224. A.R. acknowledges support from the the Swiss National Science Foundation for funding through Postdoc.Mobility (Grant No. P500PT225461). PS acknowledges support from Swiss National Science Foundation (NCCR SwissMAP). M.P.-L. acknowledges support from the Grant RYC2022-036958-I funded by MICIU/AEI/10.13039/501100011033 and by ESF+. J.~C.  is supported by Spanish MICINN AEI PID2022-141283NB-I00, and MCIN with funding from European Union NextGenerationEU (PRTR-C17.I1) and by Generalitat de Catalunya. J.~C. also acknowledges support from ICREA Academia award.

\bibliography{planckianBIB}

\newpage \
\newpage

\widetext
\begin{center}
\textbf{\large Supplemental Material for ``An information-theoretic proof of the Planckian bound for thermalization''}
\end{center}
\makeatletter
\@addtoreset{equation}{section}
\makeatother
\setcounter{figure}{0}
\setcounter{table}{0}
\setcounter{page}{1}
\setcounter{section}{0}
\makeatletter
\renewcommand{\thesection}{\Alph{section}}
\renewcommand{\thesubsection}{\roman{subsection}}
\renewcommand{\theequation}{\thesection.\arabic{equation}}
\renewcommand{\thefigure}{S\arabic{figure}}


\section{Preamble: quantum operators, Fisher information,  notation.}
\label{app:math_preamble}
In this appendix, we remind the definitions of common mathematical quantities used in the context of quantum information, as well as operators that we use in the derivation of our results.

\paragraph*{\underline{Fidelity,  Bures angle, distances and joint convexity.}}
The fidelity $F$ between two quantum states $\rho$ and $\sigma$ is given by
\begin{align}
    \label{eqapp:fidelity}
    F(\rho,\sigma)=F(\sigma,\rho):=\Tr{\sqrt{\sqrt{\sigma}\rho\sqrt{\sigma}}}^2\;,
\end{align}
and it can achieve values between $F=0$ (perfectly distinguishable, orthogonal states) and $F=1$ ($\rho=\sigma$).
The \emph{Bures angle} (or quantum Bhattacharyya angle) is given by
\begin{align}
\label{eqapp:bures_angle}
   D_A(\rho,\sigma):= \arccos(\sqrt{F(\rho,\sigma)})\;,
\end{align}
and is bounded as $0\leq D_A\leq \frac{\pi}{2}$ accordingly.
The related \emph{Bures distance} $D_B(\rho,\sigma)$ between two quantum states is defined as
\begin{align}
\label{eqapp:bures_dist}
    D_B(\rho,\sigma)^2:=2\left(1-   \sqrt{F(\rho,\sigma)}\right)\;.
\end{align}
$D=D_B,D_A$ both define a distance on the set of states and satisfy in particular the triangular inequality
\begin{align}
\label{eqapp:triang_ineq_D}
    D(\rho,\sigma)+D(\sigma,\kappa)\geq D(\rho,\kappa)
\end{align}
which is crucial to our purposes.
Moreover both the squared Bures distance and squared Bures angle are additionally \emph{jointly convex}, namely
\begin{align}
     D^2(\lambda \rho_1+(1-\lambda)\rho_2,\lambda \sigma_1+(1-\lambda)\sigma_2)\leq \lambda D^2(\rho_1,\sigma_1)+(1-\lambda)D^2(\rho_2,\sigma_2)\;.
\end{align}
While this property is known for $D_B^2$, it is slightly less known for $D_A^2$, reason for which we shortly prove it here for the reader. First, notice that the joint convexity of $D_B^2$ is due to the joint concavity property of the square root fidelity, i.e.
\begin{align}
\label{eqapp:sqrt_fid_conc}
    \sqrt{F(\lambda \rho_1+(1-\lambda)\rho_2,\lambda \sigma_1+(1-\lambda)\sigma_2)}\geq \lambda \sqrt{F(\rho_1,\sigma_1)}+(1-\lambda)\sqrt{F(\rho_2,\sigma_2)}\;.
\end{align}
Second, notice that the function $\arccos(x)^2$ is convex and decreases monotonically for $0\leq x\leq 1$. It follows that
\begin{align}
\nonumber
    \arccos^2\left(\sqrt{F(\lambda\rho_1+(1-\lambda\rho_2),\lambda \sigma_1+(1-\lambda)\sigma_2)}\right) &\leq \arccos^2\left(\lambda \sqrt{F(\rho_1,\sigma_1)}+(1-\lambda)\sqrt{F(\rho_2,\sigma_2)}\right)\\
    &\leq \lambda \arccos^2\left(\sqrt{F(\rho_1,\sigma_1)}\right)+(1-\lambda)\arccos^2\left(\sqrt{F(\rho_2,\sigma_2)}\right)\;,
\label{eqapp:D_A^2_JC_proof}
\end{align}
where the first inequality follows from~\eqref{eqapp:sqrt_fid_conc} and the decreasing monotonicity of $\arccos^2(x)$, and the second inequality from the convexity of $\arccos^2(x)$. Eq.~\eqref{eqapp:D_A^2_JC_proof} shows the joint convexity of $D_A^2$.

\paragraph*{\underline{Bures metrics and Quantum Fisher information.}}
Locally, the Bures angle and Bures distance both define the same Riemannian metric on the set of density matrices\footnote{$D_B$ represents the geodesics distance relative to~\eqref{eqapp:DBDA_metric} between $\rho$ and $\sigma$ in the manifold of positive semidefinite matrices, whereas $D_A$ corresponds to the same metrics integrated on the manifold of normalized density matrices satisfying the additional constraint $\Tr{\rho}=1$, see e.g.~\cite{scandi2023quantum}.}
\begin{align}
\label{eqapp:DBDA_metric}
    D_B(\rho,\rho+\delta\rho)^2\simeq D_A(\rho,\rho+\delta\rho)^2 &\simeq \frac{1}{4} \llangle\delta\rho|\delta\rho \rrangle|_{\rho} + \mathcal{O}(|\delta\rho|^3)
\end{align}
that is quantified by the scalar product associated to the (Bures) Quantum Fisher information (QFI), namely
\begin{align}
    \llangle\delta\rho|\delta\rho \rrangle|_{\rho}&=\Tr{\delta\rho\J^{-1}_\rho[\delta\rho]}\;,
\quad \text{where} \quad
    \J_\rho[A]=\frac{\rho A+A\rho}{2}\;.
\end{align}
$\J_\rho$ can be inverted as
\begin{align}
\label{eqapp:J_bur}
    \J_{\rho}[A] &:=\frac{\rho A+A\rho}{2}\;, &   \J^{-1}_{\rho}[A] &= 2\int_0^\infty\D s \; e^{-\rho s} A e^{-\rho s}.
\end{align}
The $\J$ operator
belongs to the family of generalized quantum Petz-Fisher multiplications~\cite{petzMonotoneMetricsMatrix1996,scandi2023quantum}  which represent non-commutative versions of the multiplication times $\rho$.
These operators are all self-adjoint w.r.t. the Hilbert-Schmidt scalar product, meaning that $\Tr{A\J[B]}=\Tr{\J[A]B}$ for Hermitian $A,B$. 
In addition to their operator-integral expressions above, notice that the action of these superoperators can be explicited in the components $\ketbra{i}{j}$, where $\ket{i}$ are the eigenvectors of $\rho$. That is, given
\begin{align}
    \rho=\sum_i p_i \ketbra{i}{i}\;, 
\end{align}
one has
\begin{align}
\label{eqapp:J_B_coordi}
    \J_{\rho}[\ketbra{i}{j}] =\frac{p_i+p_j}{2} \ketbra{i}{j}\; .
\end{align}

In general, the QFI $\F(\rho_\theta)$ of a state parametrized by some $\theta$ can be equivalently expressed as
\begin{align}
    \F(\rho_\theta):=  \llangle\partial_\theta\rho_\theta|\partial_\theta\rho_\theta \rrangle|_{\rho_\theta}&=\Tr{\partial_\theta\rho_\theta\J^{-1}_{\rho_\theta}[\partial_\theta\rho_\theta]}\equiv\lim_{\delta\theta\rightarrow 0}4 \frac{D(\rho_\theta,\rho_{\theta+\delta\theta})^2}{\delta\theta^2}\;.
    \label{eqapp:Bures_metrics}
\end{align}

\begin{center}
    \emph{\fbox{NOTE:} \\in the rest of the supplementary material and throughout the main text,\\ we simply indicate the Bures angle $D_A$~\eqref{eqapp:bures_angle} as $D$.}
\end{center}

\paragraph*{\underline{Spectral seminorm, spectral norm, and thermodynamics.}}
The spectral seminorm $\|\dots\|$ of a hermitian operator is given by the difference of the maximum and minimum eigenvalues
\begin{align}
\label{eqapp:spec_seminorm}
    \|A\|:=\lambda_{\rm max}(A)-\lambda_{\rm min}(A)\;.
\end{align}
Notice that $\|\dots\|$ is a seminorm as it satisfies homogeneity $ \|\eta A\|=\eta \|A\|$ and the norm inequality $\|A+B\|\leq \|A\|+ \|B\|$. However it can be zero for nonnull operators, namely those proportional to the identity operator, $ \|\eta\id\|\equiv 0$. 
We can define the equivalence relation between operators $A\sim B$ if $A-B\propto\id$. In such case, the quotient space of operators is still a vector space, with a null vector given by the class $0:=[\id]$. On such space, $\|\dots\|$ is a proper norm. Notice that the quotient space of Hermitian operators (bounded from below in the infinite-dimensional case) is exactly the natural space of Hamiltonians, as energy is defined up to a constant offset, and $H$ and $H+\eta\id$ define the same thermal state for any $\eta$.

\section{Finite distance and QFI for unitary evolutions, generalized Heisenberg limit, entanglement.}
\label{app:dynQFI}
Consider the reduced state of a unitary evolution, as in the main text is the system $S$, that is
\begin{align}
    \rho_S(t,H_S)={\rm Tr}_M \rho_{SM}(t) ={\rm Tr}_M \; U(t) \rho_{SM}(0) U^\dagger(t)\;,
\end{align}
where
\begin{align}
\label{eqapp:Texp}
    U(t)=\mathcal{T}\exp{\frac{i}{\hbar}\int_0^t\D s\; H_{tot}(s) }\;,\quad H_{tot}(s)=H_S+V_{SM}(s)+H_M(s)\;.
\end{align}

Our goal is to upper bound the value of the finite Bures angle between two states that evolved with different Hamiltonians $H_S^{(1)}$ and $H_S^{(2)}$.

We start by using a standard data-processing-inequality induced by the partial trace, namely
\begin{align}
\label{eqapp:Tr_DPI}
    D^2(\rho_S(t,H_S^{(1)}),\rho_S(t,H_S^{(2)})\leq D^2(\rho_{SM}(t,H_{tot}^{(1)}),\rho_{SM}(t,H_{tot}^{(2)}))\;.
\end{align}
Secondly, we consider the local metric properties of $D$. Namely, as $D(\rho,\sigma)$ corresponds to the geodesics length of the metrics~\eqref{eqapp:Bures_metrics} integrated on the set of normalized density matrices, it follows that any other trajectory with the same initial and final points will be longer,
\begin{align}
 D(\rho,\sigma)^2\leq \frac{1}{4}\int_0^1 \D s \llangle\dot\eta(s)|\dot{\eta}(s)\rrangle|_{\eta(s)}    \quad \text{for any trajectory satisfying} \quad \eta(s=0)=\rho, \;\eta(s=1)=\sigma .  
\end{align}
For our purposes we can specify this relation to trajectories of states corresponding to different evolutions induced by $H_S$ time $\tau$, that is $\eta(s)=\rho_S(\tau,H_s)$ with $H_{s=0}=H_S^{(1)}$ and $H_{s=1}=H_S^{(2)}$, from which
\begin{align}
    D^2(\rho_{SM}(\tau,H_S^{(1)}),\rho_{SM}(\tau,H_S^{(2)}))\leq\frac{1}{4} \int \D s \llangle\partial_s\rho_{SM}(\tau,H_s)|\partial_s\rho_{SM}(\tau,H_s)\rrangle_{\rho_{SM}(\tau,H_s)}\;.
\end{align}
Taking specifically $H_s=H_1+s(H_2-H_1)$ we can use the known bound~\cite{Boixo2007,paris2009quantum}
\begin{align}
    \llangle\partial_s\rho_{SM}(\tau,H_s)|\partial_s\rho_{SM}(\tau,H_s)\rrangle_{\rho_{SM}(\tau,H_s)}\leq \frac{\tau^2\|H_S^{(1)}-H_S^{(2)}\|^2}{\hbar^2}
    \label{eqapp:dyn_fish_upper}
\end{align}
to finally obtain the bound on the distance between the evolved states on the total $S+M$ space
\begin{align}
\label{eqapp:finite_SM_dyn_bound}
    D^2(\rho_{SM}(\tau,H_S^{(1)}),\rho_{SM}(\tau,H_S^{(2)}))\leq \frac{\tau^2\|H_S^{(1)}-H_S^{(2)}\|^2}{4\hbar^2}\;.
\end{align}
which bounds as well the distance on $S$ alone
\begin{align}
\label{eqapp:Dbou_S}
    D(\rho_S(\tau,H^{(1)}),\rho_S(\tau,H^{(2)}))\leq  \frac{\tau\|H_S^{(1)}-H_S^{(2)}\|}{2\hbar}\;
\end{align}
via~\eqref{eqapp:Tr_DPI}.

\subsection{Average-in-time, dynamical bound, entanglement}
\label{subsec:dynamical_Fish_time_average}
For completeness and clarity we re-derive the known bound~\eqref{eqapp:dyn_fish_upper} here for any unitarily-parametrized trajectory of states.
More precisely, consider the unitary system-machine dynamics~\eqref{eqapp:Texp}, which is of the form $\rho_\theta(t)=U_\theta(t)\rho(0)U_\theta^\dagger(t)$ where we let the local Hamiltonian $H_S(\theta)$ to be parametrized by some $\theta$. One can compute the Quantum Fisher information as
\begin{align}
    \F(\rho_\theta)=\llangle\partial_\theta\rho_\theta(t)|\partial_\theta\rho_\theta(t)\rrangle|_{\rho_\theta(t)}
\end{align}
using
\begin{align}
    \partial_\theta\rho_\theta(t)&=\partial_\theta U_\theta(t)\rho_\theta U^\dagger_\theta(t)+U_\theta(t)\rho_\theta(t)\partial_\theta U^\dagger_\theta(t)=-i\frac{t}{\hbar}[\tilde{H}_\theta(t),\rho_\theta(t)]\;,\\
    \tilde{H}_\theta(t):&=i\frac{\hbar}{t}\partial_\theta U_\theta(t)U^\dagger_\theta(t)\;.
\end{align}
The explicit expression for $\tilde{H}_\theta(t)$ can thus be computed from~\eqref{eqapp:Texp} as
\begin{align}
\label{eqapp:Htilda}
    \tilde{H}_\theta(t)=\int_0^1 \D s\, U_\theta(t,st)\dot{H}_\theta U_\theta({st,0})U_\theta^{\dagger}(t,0)=\int_0^1 \D s\,U_\theta(t,st)\dot{H}_\theta U_\theta^{\dagger}(t,st):=\int_0^1\D s \dot{H}_\theta^{(s)}\;,
\end{align}
where we indicate by $\dot{H}_\theta\equiv\partial_\theta H_\theta$ which is constant in time, and we introduced the $(t,t')$ propagator for $t\geq t'$
\begin{align}
    U(t,t')=\mathcal{T}\exp{\frac{i}{\hbar}\int_{t'}^t\D s\; H_{\theta}(s)}\;.
\end{align}
As $\tilde{H}_\theta(t)$ corresponds to an integral average~\eqref{eqapp:Htilda}, we can use the convexity of the QFI to bound
\begin{align}
    \llangle\partial_\theta\rho_\theta(t)|\partial_\theta\rho_\theta(t)\rrangle|_{\rho_\theta(t)}=\frac{t^2}{\hbar^2}\llangle[\tilde{H}_\theta(t),\rho_\theta(t)]|[\tilde{H}_\theta(t),\rho_\theta(t)]\rrangle|_{\rho_\theta(t)}\leq \int_0^1\D s \frac{t^2}{\hbar^2}\llangle[\tilde{H}_\theta^{(s)},\rho_\theta(t)]|[\tilde{H}_\theta^{(s)},\rho_\theta(t)]\rrangle|_{\rho_\theta(t)}\;.
\end{align}
The last expression can be explicitly rewritten as
\begin{align}
    \nonumber
    &\frac{t^2}{\hbar^2} \int_0^1\D s\; \Tr{[U_\theta(t,st)\dot{H}_\theta U_\theta^{\dagger}(t,st),\rho_\theta(t)]\J^{-1}_{\rho_\theta}(t)[\rho_\theta(t),U_\theta(t,st)\dot{H}_\theta U_\theta^{\dagger}(t,st)]}\\
    =&\frac{t^2}{\hbar^2}\int_0^1\D s\; \Tr{[\dot{H}_\theta,\rho_\theta(t-st)]\J^{-1}_{\rho_\theta(t-st)}[\rho_\theta(t-st),\dot{H}_\theta]}\nonumber\\
    =& \frac{t^2}{\hbar^2}\int_0^1\D s\; \Tr{[\dot{H}_\theta,\rho_\theta(st)]\J^{-1}_{\rho_\theta(st)}[\rho_\theta(st),\dot{H}_\theta]}\;.
\end{align}
We thus found a meaningful lemma: \emph{the QFI of a parametric time-dependent unitary evolution generated by $H_\theta(t)$ (with fixed derivative $\partial_\theta H_\theta(t)\equiv\partial_\theta H_\theta=:\dot{H}_\theta$, is bounded from above by the average QFI of unitary rotations generated by $\dot{H}_\theta$ on $\rho(st)$ averaged on all times $st$ smaller than $t$.}

\paragraph*{\underline{State-agnostic bound.}}
In the absence of further information the upperbound on the QFI
\begin{align}
\label{eapp:Fbou2}
    \F(\rho_\theta)\leq \frac{t^2}{\hbar^2}\int_0^1\D s\; \Tr{[\dot{H}_\theta,\rho_\theta(st)]\J^{-1}_{\rho_\theta(st)}[\rho_\theta(st),\dot{H}_\theta]}
\end{align}
can finally be bounded using again the convexity of the QFI and noticing that each term is bounded the pure state QFI, that is the variance of $\dot{H}_\theta$~\cite{paris2009quantum}
\begin{align}
    \Tr{[\dot{H}_\theta,\rho_\theta(st)]\J^{-1}_{\rho_\theta(st)}[\rho_\theta(st),\dot{H}_\theta]}\leq \sum_k p_k\bra{\psi_k}\Delta^2\dot{H}^2_\theta\ket{\psi_k}
\end{align}
where $\rho_\theta(st)=\sum_k\ketbra{\psi_k}{\psi_k}$.
Finally, the variance of any operator $O$ is bounded by $\|O\|^2/4$, leading to Eq.~\eqref{eqapp:finite_SM_dyn_bound}.

\paragraph*{\underline{A comment on entanglement}}
Consider now the case of $n$-partite Hamiltonian acting on $S$, that is of the form 
\begin{align}
    \partial_\theta H_\theta=\sum_{i=1}^n h_i
\end{align}
in which each $h_i\equiv\id_1\otimes\id_2\otimes\dots\otimes h_i\otimes \id_{i+1}\otimes\dots\otimes\id_n\otimes \id_M$ acts locally on the $i$-th site of the system $S$, and trivially on $M$.
Suppose now that $\rho_\theta$ in~\eqref{eapp:Fbou2} is $k$-separable, meaning that it can be written as the convex sum of states with at most $k$-partite entanglement on $S$,
\begin{align}
    \rho_\theta=\sum_\alpha \rho^{(\alpha_1)}_{SM_{\alpha_1}}\otimes \rho^{(\alpha_2)}_{SM_{\alpha_2}}\otimes\dots\;.
\end{align}
Here, $\alpha$ is a label and $SM_{\alpha_i}$ is the $i$-th set of a partition of the total $S+M$ system, containing at most $k$ sites of $S$.
It follows then that the QFI $\F_\theta(\rho_\theta)$ is bounded, via convexity, to
\begin{align}
    \frac{\hbar^2}{t^2}\F_\theta(\rho_\theta)\leq\sum_{\vec{\alpha}} p_{\alpha_1^{(i_1)},\alpha_2^{(i_2)},\dots}\bra{\psi_{\alpha_1^{(i_1)}}}\bra{\psi_{\alpha_2^{(i_2)}}}\dots \Delta^2\left(\sum_i h_i\right) \ket{\psi_{\alpha_1^{(i_1)}}}\ket{\psi_{\alpha_2^{(i_2)}}}\dots
\end{align}
where the $\ket{\psi_{\alpha_j}^{(i_j)}}$ are the eigenvectors of the states $\rho_{\alpha_j}$, and $p_{\alpha_1^{(i_1)},\alpha_2^{(i_2)},\dots}$ the corresponding probabilities.
From convexity and the summation property of the variance for factorized states it then follows the bound
\begin{align}
     \frac{\hbar^2}{t^2}\F_\theta(\rho_\theta)\leq \sum_{\vec{\alpha}} p_{\alpha_1,\alpha_2,\dots} \frac{1}{4}\left( \| \sum_{i\in{S_{\alpha_1}}} h_i\|^2+\|\sum_{i\in{S_{\alpha_2}}} h_i\|^2+\dots\right)\;. 
\end{align}
Finally, we notice that this last quantity scales at most as $k^2 \cdot\lceil\frac{n}{k}\rceil$, where $k$ is the maximum size (in sites) of the sets $S_{\alpha_i}$.
In particular, if all the $h_i$ have the same spectral seminorm $\|h_i\|\equiv\|h\|$ $\forall i$ (as, for example, in the case of a uniform field), we find
\begin{align}
    \F_\theta(\rho_\theta)\leq \frac{t^2}{\hbar^2}k^2 \lceil\frac{n}{k}\rceil\frac{\|h\|^2}{4}
\end{align}
which is valid for $k$-separable $\rho_\theta$.

\section{General framework and Planckian bound derivation}
\label{app:formal_derivation}
In this appendix we provide a complete formal description of the thermalization framework used in our work, and the following derivation of the general Planckian bound~\eqref{eq:chi_def} on thermalization time.

In~\ref{subsec:weakest_therm} we formalize the notion of (quantum) thermalization that we use in the main text -- specifically, requirement~\ref{ass:local_term} -- and include the possibility of a finite time-resolution in the observation of the thermal state.

In~\ref{subsec:general_derivation} we provide the derivation of our main bound~\eqref{eq:chi_def}.

\subsection{A weak nontrivial notion of thermalization}
\label{subsec:weakest_therm}

We expect thermalization to be a dynamical mechanism bringing a system close to its thermal ensemble, when in contact with a much larger environment. 
That is, at some time $\tau$ of the dynamics, one should have
\begin{align}
    \rho_S(\tau,H_S)\approx \frac{e^{-\beta H_S}}{\Tr{e^{-\beta H_S}}}=:\omega(\beta,H_S)
    \label{eq:intuitive_therm}
\end{align}
in some mathematically well-defined sense, for the proposed temperature $T=(k_B\beta)^{-1}$.

First of all, we require the dynamics to be described by standard nonrelativistic quantum mechanics (Requirement~\ref{ass:q_mech}), that is for some dilation of the system+machine the dynamics is closed and unitary, possibly time-dependent, i.e. governed by the Schrödinger equation
\begin{align}
    \frac{d}{dt}\rho_{SM}=-\frac{i}{\hbar}[H_S+V_{SM}(t)+H_M(t),\rho_{SM}(t)]
\end{align}
from which the dynamics can be integrated as~\eqref{eqapp:Texp}
\begin{align}
    \rho_{SM}(t)&=U(t)\rho_{SM}(0)U^\dagger(t)\;,\\
    \rho_{S}(t)&={\rm Tr}_M \rho_{SM}(t)\;.
\end{align}
Two remarks here:

1) Notice that the degrees of freedom of the environment/machine $M$ might include also d.o.f. of the same particle, for example, the position $x$ of a spin-$\frac{1}{2}$ particle, that only thermalizes in the $2$-dimensional space of the spin d.o.f..

2) Notice as well that in standard setups one could have an uncorrelated initial state of the form $\rho_{SM}(0)=\rho_S(0)\otimes\rho_M(0)$, however such assumption is not necessary to our purposes.

We now turn to formalizing the notion of thermalization. We make the weakest possible assumptions on the specifics of Eq.~\eqref{eq:intuitive_therm}: 
\begin{enumerate}    
    \item \underline{Nontrivial local  thermalization}: in order to have a proper thermalization process\footnote{As discussed in the main text, thermalization for a single value of the local $H_S$ can be trivially obtained by storing copies of $\omega(\beta,H_S)$ in advance.}, $\eqref{eq:intuitive_therm}$ should be valid at least for more than one single Hamiltonian. In order to derive a nontrivial Planckian bound, it will be enough to consider $H_S$ taking two values, $H_S^{(1)}$ and $H_S^{(2)}$ (cf.~\ref{subsec:general_derivation}). However, for simplicity we will often consider in some continuous set of Hamiltonian $\mathcal{H}$, such as the neighborhood $\mathcal{H}\equiv\mathcal{B}_\delta$ of a given $\bar{H}_S$
    \begin{align}
        \mathcal{B}_\delta:\{H_S \big|\; \|H_S-\Bar{H}_S\|\leq \delta\}
    \end{align}
    according to some norm. 
    \item \underline{Thermalization is approximate}: for all values of $H_S\in\mathcal{H}$,  one should have 
    \begin{align}
        D(\rho_S(\tau,H_S),\omega(\beta,H_S))\leq \varepsilon \quad \forall H_S\in\mathcal{H}\;,
    \end{align}
    according to some notion of distance $D$. Here $\varepsilon$ thus indicates the admitted error of the thermalization process.
    \item \underline{Time-resolution}: finally we also allow for a non-exact time-resolution of the measurement devices. That is, we assume that rather than $\rho_S(\tau,H)$, the state sampled by the machine's user is given by a time-average 
    \begin{align}
    \label{eqapp:timeres_mu}
        \tilde{\rho}_S(\tau,H_S):=\int_{0}^{\tau}\D t\;\mu(t)\rho_S(t,H_S) \quad \text{for some measure } \int_0^\tau\D t\;\mu(t)=1\;.
    \end{align}
\end{enumerate}
These conditions correspond to (a more general version of) machine requirement~\ref{ass:local_term} in the main text.

\paragraph*{Definition of thermalization.} Putting together the conditions above, we arrive to the formal statement of what is thermalization to the canonical ensemble, in its weakest form, that is we require
\begin{align}
        D(\tilde{\rho}_S(\tau,H_S),\omega_\beta(H_S))\leq \varepsilon\quad \forall   H_S\in\mathcal{H}\;. 
        \label{eqapp:formal_thermalization}
\end{align}


\subsection{Formal bound}
\label{subsec:general_derivation}
\begin{figure}
    \centering
    \includegraphics[width=0.5\linewidth]{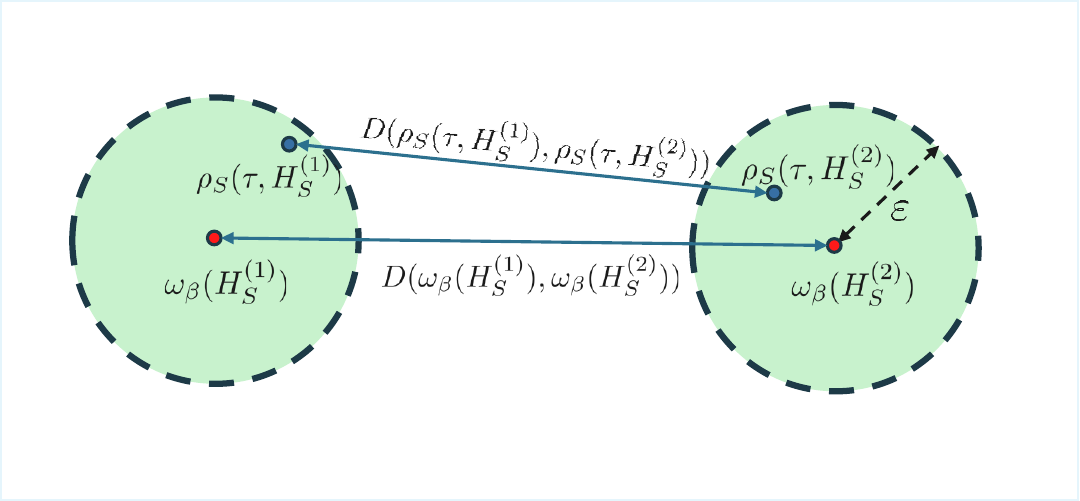}
    \caption{Illustration of the triangular inequality~\eqref{eqapp:triangle_ineq}. As the dynamics $\rho_S(\tau,H_S)$ is required to follow the local thermal states up to error $\varepsilon$, the distance between $\rho_S(\tau,H_S^{(1)})$ and $\rho_S(\tau,H_S^{(2)})$ is bounded from below.}
    \label{fig:triangle_inequality}
\end{figure}

The Planckian bound on thermalization is the consequence of a triangular inequality associated to~\eqref{eqapp:formal_thermalization}, namely for any two values $H_S^{(1)}$ and $H_S^{(2)}$ in the set of thermalizing Hamiltonians $\mathcal{H}$, it follows from~\eqref{eqapp:formal_thermalization} that (see Fig.~\ref{fig:triangle_inequality})
\begin{align}
    D(\tilde{\rho}_{S}(\tau,H_S^{(1)}),\tilde{\rho}_{S}(\tau,H_S^{(2)}))\geq D(\omega(\beta,H_S^{(1)}),\omega(\beta,H_S^{(2)}))-2\varepsilon\;.
    \label{eqapp:triangle_ineq}
\end{align}
This is the starting point of our derivation. We now pass to upper-bounding the left-hand side of~\eqref{eqapp:triangle_ineq}.
If the chosen distance $D$ features a joint-convexity property, we can use it to take into account the time-resolution of the measurement devices~\eqref{eqapp:timeres_mu}, that is, from joint convexity of $D^2$ it follows
\begin{align}
    D^2(\tilde{\rho}_{S}(t,H_1),\tilde{\rho}_{S}(t,H_2))\leq \max_{0 \leq t\leq \tau} D^2({\rho}_{S}(t,H_S^{(1)}),{\rho}_{S}(t,H_S^{(2)}))
    \label{eqapp:t'_JC}
\end{align}
Finally, we use a bound of the form (as is the case~\eqref{eqapp:Dbou_S} with our choice of $D$ and $\|\dots\|$)
\begin{align}
    D^2({\rho}_{S}(t,H_S^{(1)}),{\rho}_{S}(t,H_S^{(2)}))\leq \frac{t^2\|H_1-H_2\|^2}{4\hbar^{2}}\;.
    \label{eqapp:normbound}
\end{align}
for the norm $\|\dots\|$.It then follows by concatenating \eqref{eqapp:triangle_ineq}-\eqref{eqapp:t'_JC}-\eqref{eqapp:normbound}, and noticing $t\leq \tau$, that
\begin{align}
    \tau \geq \hbar \frac{2D(\omega_\beta(H_S^{(1)}),\omega_\beta(H_S^{(2)}))-4\varepsilon}{\|H_S^{(1)}-H_S^{2)}\|}= \beta \hbar \frac{2D(\omega_\beta(H_S^{(1)}),\omega_\beta(H_S^{(2)}))-4\varepsilon}{\beta\|H_S^{(1)}-H_S^{(2)}\|}\;.
\end{align}
This bound needs to be satisfied in the whole range of Hamiltonians for which we request the thermalization to be valid, i.e. $\forall H_S^{(1)},H_S^{(2)}\in\mathcal{H}$. 
We therefore have in general the relation
\begin{align}
    \tau \geq \beta\hbar\;\chi(\mathcal{H},\varepsilon)\;,
\end{align}
where
\begin{align}
\label{eqapp:chi_def}
    \chi(\mathcal{H},\varepsilon)=\max_{H_S^{(1)},H_S^{(2)}\in \mathcal{H}} \frac{2D(\omega_\beta(H_S^{(1)}),\omega_\beta(H_S^{(2)}))-4\varepsilon}{\beta\|H_S^{(1)}-H_S^{(2)}\|}\;.
\end{align}
This is our main result, presented also in the main text~\eqref{eq:chi_def}.
A few remarks are in order: first, notice that $\chi$ solely depends on $\beta H_S$ rather than $H_S$ itself. This makes it dimensionless and, in general, (as we prove below) finite. More specifically, $\chi$ can be seen to depend on the set ${\Omega}$ spanned by the thermal states obtained from Hamiltonians in $\mathcal{H}$ (at temperature $\beta^{-1}$).
Moreover, notice that in general for the bound to be nontrivial, $2\varepsilon$ should be smaller than $D(\omega(\beta,H^{(1)}_S),\omega(\beta,H^{(2)}_S)$ for some choice of thermal states in the region $\mathcal{H}$ under scrutiny.

In this work, we choose $D\equiv D_A$ to be the Bures angle~\eqref{eqapp:bures_angle}, and $\|\dots\|$ the spectral seminorm~\eqref{eqapp:spec_seminorm},
for which it holds the upperbound~\eqref{eqapp:Dbou_S}.
Notice that with this choice of normalization, the maximum distance between two states is given by $\pi/2$, therefore the tolerated $\varepsilon$ should be between $0$ and 
\begin{align}
    \varepsilon_{\rm max}=\frac{\pi}{4}\;
\end{align}
in order to yield a nontrivial bound. In fact, for any two given (thermal) states $\omega(\beta,H_S^{(1)})$ and  $\omega(\beta,H_S^{(2)})$, it is possible to  find a state $\bar{\omega}$ that satisfies $D(\omega(\beta,H_S^{(1)}),\bar\omega)\leq \pi/4$ and $D(\omega(\beta,H_S^{(2)}),\bar\omega)\leq \pi/4$. The instantaneous preparation of $\bar\omega$ would then yield a $\varepsilon_{\rm max}$-thermalization machine for $\mathcal{H}\equiv\{H_S^{(1)},H_S^{(2)}\}$ (as well as for all the Hamiltonians generating thermal states $\varepsilon_{\rm max}$-close to $\bar{\omega}$). In the qubit case, the maximally mixed state $\id/2$ satisfies $D(\id/2,\rho)\leq\varepsilon_{\rm max}$ $\forall\rho$.

\section{Lowerbounds to $\tilde{\chi}(\rho)$ (locally-exact thermalization)}
\label{sec:fisherbounds}
The QFI of states in exponential forms 
\begin{align}
    \rho_\theta=\frac{e^{-{X_\theta}}}{\Tr{e^{-{X_\theta}}}}
\end{align}
(such as Gibbs states in natural units where $X=\beta H$), is given by substituting the exponential derivative
\begin{align}
    \partial_\theta\rho_\theta=\left(\int_0^1 \D s\frac{e^{-sX_\theta}\partial_\theta X_\theta e^{-(1-s)X_\theta}}{\Tr{e^{-X_\theta}}}\right)- \frac{e^{-{X_\theta}}}{\Tr{e^{-{X_\theta}}}}\Tr{\partial_\theta X_\theta \frac{e^{-{X_\theta}}}{\Tr{e^{-{X_\theta}}}}}
\end{align}
inside~\eqref{eqapp:Bures_metrics}. The resulting expression can be written, in the basis of eigenstates of $\rho_\theta=\sum_i p_{\theta_i}\ketbra{\psi_{\theta_i}}{\psi_{\theta_i}}$ as (cf.~\cite{abiuso2025fundamental})
\begin{align}
    \F_\theta=\sum_{ij} \frac{2(p_i-p_j)^2}{(\ln p_i-\ln p_j)^2(p_i+p_j)} X'_{ij} -\left(\sum_i p_i X'_{ii}\right)^2\quad \text{where} \quad X'_{ij}:=\bra{\psi_{\theta_i}}\partial_\theta X_\theta\ket{\psi_{\theta_j}}\;.
\end{align}
In the limit of locally-exact thermalization described in~\ref{sec:fisher_limit}, the adimensional factor $\tilde{\chi}$ in the Planckian bound $\tau\geq \tau_{\rm Pl}\cdot \tilde{\chi}$ simply becomes the maximisation of $\sqrt{\F_\theta}$ over all possible directions of perturbation $X'$ with spectral norm $\|X'\|=1$. We thus want to estimate
\begin{align}
\nonumber 
    \tilde{\chi}(\rho):= \max_{\|X'\|=1} \sqrt{
   \sum_i p_i {X'}_{ii}^2 -(\sum_i p_i {X'}_{ii})^2 + \sum_{ij} |X'_{ij}|^2 \frac{2(p_i-p_j)^2}{(\ln p_i-\ln p_j)^2(p_i+p_j)}
   }\;.
\end{align}
In the following, we provide several lower bounds to $\tilde{\chi}$ based on different possible choices of $X'$.

{\bf (i) Diagonal bound: }
When considering only diagonal perturbations ($X'_{ij}=0$ (for $i\neq j$)) it is relatively straightforward to maximise the corresponding classical variance 
and obtain
\begin{align}
    \tilde{\chi}(\rho)\geq \tilde{\chi}(\rho)^{\rm clas}=\sqrt{p^*(1-p^*)}
    \label{eqapp:chi_pstar}
\end{align}
which is obtained choosing e.g. $X_{ii}=1$ for a subset $i\in{I}$ of the microstates in the diagonal decomposition of $\rho$, having total probability $P({I})=p^*$, ($X_{ii}=0$ for $i\notin \mathcal{I}$ to guarantee $\|X\|$=1). That is, for any to coarse grainining of the thermal state's populations in
$\{p^*,1-p^*\}$ one has a corresponding inequality~\eqref{eqapp:chi_pstar}. Clearly the value of the coarse-grained probability that is closest to $1/2$ will yield the best bound.

For example, notice that whenever the ground state probability $p_0$ is below $p_{0}\leq \frac{2}{3}$, one can always find a coarse graining such that $\frac{1}{3}\leq p^*\leq \frac{2}{3}$, and therefore $\chi\geq \frac{\sqrt{2}}{3}\sim 0.47$.

{\bf (ii) Coherent bound: }
One can consider a different subset of (non diagonal) perturbations of the form $X'_{ij}=\frac{1}{2}$ (again guaranteeing $\|X'\|=1$) for any set of non-overlapping pairs of indices, obtaining
\begin{align}
    \tilde{\chi}^2(\rho)\geq \sum_{{\rm couples} \{ij\}} \frac{1}{4} \frac{4(p_i-p_j)^2}{(\ln p_i-\ln p_j)^2(p_i+p_j)}
\end{align}
Notice that for fixed ``fraction of population" used, one should privilege to couple probabilities that are close to each other. In the limit of very-close populations each term tends to $\frac{1}{2}p_i$. Therefore the maximum is again for $p_i=1/2$ and is the same as the classical maximum above. However, for large gaps/ very pure states it is useful to see that
\begin{align}
\label{eqapp:qubitbound}
    \sum_{{\rm couples} \{ij\}} \frac{1}{4} \frac{4(p_i-p_j)^2}{(\ln p_i-\ln p_j)^2(p_i+p_j)}\geq \frac{(p_0-p_1)^2}{(\ln p_0 -\ln p_1)^2}\geq \frac{(2p_0-1)^2}{(\beta\Delta)^2}\geq \frac{(2p_0-1)^2}{(\ln p_0 -\ln(1-p_0)+\ln (d-1))^2}\;.
\end{align}

\begin{figure}
    \centering
    \includegraphics[width=0.5\linewidth]{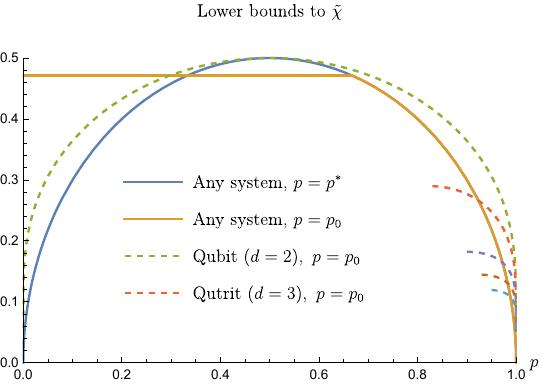}
    \caption{
    {\bf Bounds for locally-exact thermalization.} In the limit locally-exact thermalization (see text), our main bound~\eqref{eq:chi_def} reduces to $\tau\geq\tau_{\rm Pl}\cdot\tilde{\chi}$~\eqref{eq:boundQFI}, which only depends on the local thermal state considered. In the plot, different lower bounds to $\tilde{\chi}$ are shown. We find that $\tilde{\chi}\geq \sqrt{p^*(1-p^*)}$ (blue line) for all possible coarse-grainings of the thermal states' populations in two sets with total probabilities $\{p^*,1-p^*\}$. The maximum of such function is obtained for $p^*=1/2$ which is easily approximated in noncritical large systems. Moreover this immediately yields $\tilde{\chi}\geq \sqrt{2}/3\sim 0.47$ for any thermal state with ground state probability $p_0$ smaller than $p_0\leq 2/3$ (yellow line). 
    In terms of spectral gap $\Delta$, notice that ground state probability $p_0$ always satisfies $p_0\leq \frac{1}{1+e^{-\beta \Delta}}$, thus $\tilde{\chi}\geq \sqrt{e}/(1+e)\sim 0.44$ for $\beta\Delta\leq 1$. For very large values of $p_0$, it is possible to show that $\tilde{\chi}\geq (2p_0-1)/(\beta\Delta)\geq (2p_0-1)/(\ln ((d-1)p_0/(1-p_0)))$, which yields the dashed lines in the plot (for increasing dimension, in the plot $d=2,3,10,30,100$). A full analytical optimization of $\tilde\chi$ is possible for qubits (green dashed line): given $X'=\frac{1}{2}(\cos(\theta)\sigma_z+\sin(\theta)\sigma_x)$, the transversal bound $\sin\theta=1$ is optimal and always larger than the diagonal one, cf.~\eqref{eqapp:qubitbound}.}
    \label{fig:plot_fish_qubit}
\end{figure}

{\bf (iii) Full Qubit bound: }
In the case of a single Qubit one solve the maximisation exactly and interpolate between the previous two bounds by choosing up to symmetry in the Bloch-sphere $X'=\frac{1}{2}(\cos(\theta)\sigma_z+\sin(\theta)\sigma_x)$ and obtain
\begin{align}
    \tilde{\chi}^2(\rho)^{\rm Qbit}=\max_\theta  \left(\cos(\theta)^2 p(1-p)+\sin(\theta)^2 \frac{(2p-1)^2}{(\ln p -\ln (1-p))^2}\right)\;.
\end{align}
We see clearly that the second addend is always bigger than the previous, therefore $\sin(\theta)^2=1$ is preferable, yielding the dashed green line bound in Fig.~\ref{fig:plot_fish_qubit}.

\section{Lower bounds to $\chi(\mathcal{H},\varepsilon)$ (finite-size/finite-error thermalization)}
\label{sec:finite_epsilon}
In this section we provide bounds to $\chi(\mathcal{H},\varepsilon)$~\eqref{eqapp:chi_def} for finite sizes of error and Hamiltonian set $\mathcal{H}$ for which the approximate thermalization holds. In the main text and in the following we take for simplicity $\mathcal{H}$ to be a ball $\mathcal{B}_\delta$ of Hamiltonians centered in some $\bar{H}_S$ and with spectral norm radius $\delta$, thus $\chi(\mathcal{H},\varepsilon)\equiv\chi(\bar{H}_S,\delta,\varepsilon)$.
Physically speaking, as opposed to the limit $\varepsilon\ll\delta\rightarrow 0$ (infinitely precise and infinitely local thermalization), here we are crucially considering the regime in which the error $\varepsilon$ is strictly finite, and as such thermalization is also required to be valid in a finite region of Hamiltonians/thermal states. 
As noticed above, the adimensional quantity $\chi$ only depends on the thermal state $\omega(\beta,H_S)$ statistics and the adimensional exponent $\beta H_S$. For this reason we will simplify the notation in the following and consider temperature units in which $\beta=1$.

We consider Hamiltonians contained in a ball $\mathcal{B}_\delta:\{H_S| \|H_S-\bar{H}_S\|\leq \delta\}$ (clearly, for any two Hamiltonians in this set we have $\|H_S^{(1)}-H_S^{(2)}\|\leq 2\delta$).
To get a lowerbound to $\chi(\bar{H}_S,\delta,\varepsilon)$, we take the subset ansatz of the form
\begin{align}
\label{eqapp:H_kappa_ansatz}
    H^{(1)}_S=\bar{H}_S+\alpha\kappa\;, \quad H^{(2)}_S=\bar{H}_S-(1-\alpha)\kappa\;,
\end{align}
where
\begin{align}
    0\leq \alpha\leq 1\;, \quad \|\kappa\|=\delta\;,
\end{align}
and more in particular 
we consider $\kappa$ to be diagonal in the basis of $\bar{H}_S$, with eigenvalues $k_i$ and of the form
\begin{align}
\nonumber
    \kappa_i &=0 \; \forall i\in I_0\;,\\
    \kappa_i &=\delta \; \forall i\in I_1\;.
\label{eqapp:kappa_form}
\end{align}
That is $\kappa$ has only two possible eigenvalues, it represents a shift of a subset ($I_1$) of the eigen-energies $E_i$ of $\bar{H}_S$. Moreover the choice~\eqref{eqapp:H_kappa_ansatz} guarantees that both $H^{(1)}_S$ and $H^{(2)}_S$ are in $\mathcal{B}_\delta$ and $\kappa\equiv H^{(1)}_S-H^{(2)}_S$.

With such choice it is possible to compute the value of the fidelity~\eqref{eqapp:fidelity} entering the definition of $\chi$~\eqref{eqapp:chi_def} via the Bures angle $D(\rho,\sigma)=\arccos{\sqrt{F(\sigma,\rho)}}$, that is
\begin{align}
    \sqrt{F(\omega^{(1)},\omega^{(2)})}&=\sum_i \frac{e^{-(E_i+\alpha\kappa_i)/2}}{\sqrt{\sum_j e^{-(E_j-(1-\alpha)\kappa_j)}}} \frac{e^{-(E_i+\alpha\kappa_i)/2}}{\sqrt{\sum_j e^{-(E_j-(1-\alpha)\kappa_j)}}}\\
    &= \frac{q_0+q_1e^{(1-2\alpha) \delta/2}}{\sqrt{q_0+q_1 e^{-\alpha\delta }}\sqrt{q_0+q_1 e^{(1-\alpha)\delta}}}
\end{align}
where to compute the second line we defined
\begin{align}
    q_0=\sum_{i\in I_0} e^{-E_i}\;, \quad q_1=\sum_{i\in I_1} e^{-E_i}\;.
\end{align}
This allows to re-express the fidelity in terms of the coarse-grained probabilities of the thermal state of $\bar{H}_S$, that is given $p^*=q_0/(q_0+q_1)$ and $1-p^*=q_1/(q_0+q_1)$, namely
\begin{align}
    \sqrt{F(\omega^{(1)},\omega^{(2)})}
    &= \frac{p^*+(1-p^*)e^{(1-2\alpha) \delta/2}}{\sqrt{p^*+(1-p^*) e^{-\alpha\delta }}\sqrt{p^*+(1-p^*) e^{(1-\alpha)\delta}}}\;,
\end{align}
We thus find a continuous familty of bounds for $\mathcal{H}=\mathcal{B}_\delta$ as a function of all possible coarse-grainings $p^*$ of the populations of $\omega(\beta,\bar{H}_S)$  
\begin{align}
\label{eqapp:LB_chi_general}
    \chi(\mathcal{B_\delta}(\omega(\beta,\bar{H}_S)),\varepsilon)\geq 
    \frac{1}{\delta}\left(2\arccos\left(\frac{p^*+(1-p^*)e^{(1-2\alpha) \delta/2}}{\sqrt{p^*+(1-p^*) e^{-\alpha\delta }}\sqrt{p^*+(1-p^*) e^{(1-\alpha)\delta}}}\right)-4\varepsilon\right)\;.
\end{align}
\begin{figure}
    \centering
    \includegraphics[width=0.5\linewidth]{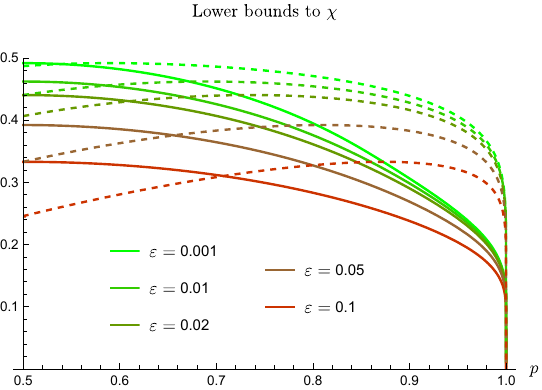}
    \caption{Lower bounds to $\chi(\bar{H}_S,\varepsilon)$ obtained by considering set of ansatz Hamiltonians~\eqref{eqapp:H_kappa_ansatz} belonging to $\mathcal{B}_\delta$. Lower bounds to $\chi$ are given as a function of the coearse-grained probability $p=p^*$ in the thermal state defined by $\bar{H}_S$, by maximising over $\delta$ the expression~\eqref{eqapp:LB_chi_general}. In particular we consider "radial variations" (dashed lines, $\alpha=0$ or $\alpha=1$), and "diametral variations" (continuous lines, $\alpha=1/2$) around $\bar{H}_S$.}
    \label{fig:pepsi_dr}
\end{figure}
In Figure~\ref{fig:pepsi_dr} we plot the bounds attainable by choosing specific $(\alpha,\delta)$-configurations described in the following.

\paragraph{Radial variations around $\bar{H}_S$.} 
Radial variations around $\bar{H}_S$  correspond to the choice $\alpha=1$ (or $\alpha=0$), that is when either $H_S^{(1)}$ or $H_S^{(2)}$ coincide with the center of the ball $\mathcal{B}_\delta$.

\paragraph{Diametral variations around $\bar{H}$.}
Diametral variations around $\bar{H}_S$  correspond to the choice $\alpha=1/2$, that is when either $H_S^{(1)}-\bar{H}_S=-(H_S^{(2)}-\bar{H}_S)$. That is such variations are symmetric around the center of the ball $\mathcal{B}_{\rm \delta}$.


\paragraph{Interpolation between the radial and diametral case.}
Numerical results indicate that diametral perturbations ($\alpha=1/2$) yield the best lower bounds for systems in which it is possible to find $p^*\approx 1/2$, whereas radial perturbations ($\alpha=0,1$) provide better bounds in the (ground state) limit $p^*\rightarrow 1$. Intuitively, this reflects the fact that close to mixed states the QFI (and its Bures angle integral) is maximal and isotropic, while for thermal states close to being pure ($\beta\Delta\gg 1$) there is no sensitivity in increasing the energy of the excited levels, as the state is already close to its ground microstate (the susceptibility to such energy variations goes to zero exponentially), whereas increasing the energy of the ground state allows a nontrivial variation of the thermal state.

In the main text Fig.~\ref{fig:main_lowerbounds}, we plot the numerical maximisation of~\eqref{eqapp:LB_chi_general} for given $p^*$  and $\varepsilon$ over $\alpha$ and $\delta$, which interpolates between radial and diametral perturbations in general, providing the highest lower bound in each case.

\begin{figure}
    \centering
    \includegraphics[width=0.6\linewidth]{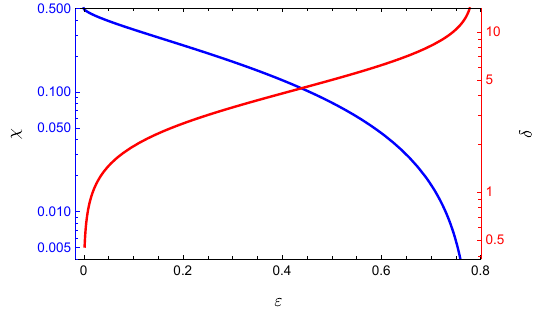}
    \caption{{\bf Accuracy-generality tradeoff.} For a thermal state in which a 50/50 population split ($p^*=0.5$) can be identified the $\pm\delta$ energy shift offers lowerbounds to $\chi$ that smoothly interpolate between $\chi\geq 0.5$ at zero error (only infinitesimal $\delta$ required, namely, locally-exact thermalization) and progressively smaller lower bounds at larger errors ($\chi\geq 0$ only in the limit of 100\% thermalization error, namely $\varepsilon=\varepsilon_{\rm max}=\pi/4$).}
    \label{fig:accuracy-generality}
\end{figure}

\paragraph{Accuracy-generality tradeoff}
As discussed above, when introducing the possibility of a finite error $\varepsilon$ in the thermalization process, the size of the set $\mathcal{H}$ of thermalizing Hamiltonian needs to be sufficiently large in order to provide
In Fig.~\ref{fig:accuracy-generality} we plot the lower bound~\eqref{eqapp:LB_chi_general} obtained for $p^*=0.5$ and optimized over $\delta$ for each value of $0\leq\varepsilon\leq \varepsilon_{\rm max}$. We notice how it smoothly interpolates between $\chi\geq 1/2$ (for $\epsilon=0$) and $\chi\geq 0$ for $\epsilon =\pi/2$. 
For each error we also plot the corresponding $\delta$ needed to obtain the lower bound. As expected, as bigger errors are tolerated by the user, more distinguishable thermal states should be asked to the machine.

\section{The optimal thermalization machine for two Hamiltonians}

\label{app: opt two}

We now consider machines that thermalize, i.e. prepare Gibbs states $\rho_S:=\omega(\beta, H_S^{(1)})$ and $\sigma_S:=\omega(\beta, H_S^{(2)})$ of, two fixed but unknown Hamiltonians $H_S^{(1)}$ or $H_S^{(2)}$. \\

By Uhlmann's theorem \cite{nielsen2010quantum}, the two states $\rho_S$ and $\sigma_S$ admit purifications $\ket{\Psi_\rho}_{SM}$ and $\ket{\Psi_\sigma}_{SM}$ on an extended system $SM$, i.e. ${\rm Tr}_M \ketbra{\Psi_\rho}_{SM}= \rho_S$ and  ${\rm Tr}_M  \ketbra{\Psi_\sigma}_{SM}= \sigma_S$, such that  
\begin{equation}
|\braket{\Psi_\rho}{\Psi_\sigma}|^2 = F(\rho,\sigma) \quad\Longleftrightarrow \quad\arccos |\braket{\Psi_\rho}{\Psi_\sigma}| = D(\rho,\sigma).
\end{equation}
By definition, a machine preparing the purifications $\ket{\Psi_\rho}$, $\ket{\Psi_\sigma}$ fulfills the thermalization task.\\

Consider a simple machine (achieving optimal Hamiltonian discrimination~\cite{aharonov2002measuring}) with sets the ``interaction" term to $V_{SM}=V_{S}=-H_S^{(1)}$, and prepares the system in the initial state 
\begin{equation}
    \ket{\psi(0)}_S= \frac{1}{\sqrt{2}} (\ket{\uparrow} +\ket{\downarrow})_S
\end{equation}
where $\ket{\uparrow(\downarrow)}$ are the eigenstates of  of $\kappa=H_S^{(2)}-H_S^{(1)}$ corresponding to its maximal (minimal) eigenvalues $\lambda_{\uparrow(\downarrow)}$.
After time $\tau$ it prepares the system in the two possible states 
\begin{align}
    \ket{\psi_1(\tau)} &= e^{-i \frac{\tau}{\hbar}(H_S^{(1)}-H_S^{(1)})}  \ket{\psi(0)}= \frac{1}{\sqrt{2}} (\ket{\uparrow} +\ket{\downarrow})\\
    \ket{\psi_2(\tau)} &=  \frac{1}{\sqrt{2}} e^{-i \frac{\tau}{\hbar}(H_S^{(2)}-H_S^{(1)})}(\ket{\uparrow} +\ket{\downarrow}) = \frac{1}{\sqrt{2}} (e^{-i \frac{\tau}{\hbar} \lambda_\uparrow}\ket{\uparrow} +e^{-i \frac{\tau}{\hbar} \lambda_\downarrow}\ket{\downarrow}) 
\end{align}
corresponding to the tow possible value system Hamiltonians $H_S^{(1)}$ and $H_S^{(2)}$. The overlap of the two states is given by 
\begin{equation}
    |\braket{\psi_1(\tau)}{\psi_2(\tau)}|^2 = \left|\frac{e^{-i \frac{\tau}{\hbar}\lambda_\uparrow}+e^{-i \frac{\tau}{\hbar} \lambda_\downarrow}}{2}\right|^2 = \cos ^2\left(\frac{\|\kappa\|  \tau }{2 \hbar}\right).
\end{equation}

Finally, note that by an isometry the two state $\ket{\psi_1(\tau)}_S$ and $\ket{\psi_2(\tau)}_S$ can be mapped (up to an irrelevant phase) to another pair of states
\begin{align}
    U \ket{\psi_1(\tau)}_S\ket{0}_M &= \ket{\Psi_\rho}_{SA}\\
    U \ket{\psi_2(\tau)}_S\ket{0}_M &= \ket{\Psi_\sigma}_{SA}
\end{align}
if only and only if the pairs of states have the same overlap $|\braket{\psi_1}{\psi_2}|=|\braket{\Psi_\sigma}{\Psi_\rho}|=\sqrt{F}$. To see this formally, note that we can always write
\begin{align}
    \ket{\psi_1(\tau)}_S &= \ket{1}_S \,\,\,\,\qquad e^{i  \phi} \ket{\psi_2(\tau)}_S =  \sqrt{F} \ket{1}_S + \sqrt{1-F} \ket{2}_S \\
     \ket{\Psi_\rho}_{SM} &= \ket{\tilde 1}_{SM} \qquad e^{i \phi'} \ket{\Psi_\sigma}_{SM} =  \sqrt{F} \ket{\tilde 1}_{SM} + \sqrt{1-F} \ket{\tilde 2}_{SM}
\end{align}
with orthogonal states $\braket{1}{2}=\braket{\tilde 1}{ \tilde 2}=0$. Now, the isometry $V = \ketbra{\tilde 0}{0} + \ketbra{\tilde 1}{1}$ gives
\begin{align}
    V \ket{\psi_1(\tau)}_S &= \ket{\Psi_\rho}_{SM} \\
    V \ket{\psi_2(\tau)}_S &= e^{-i \phi}V (\sqrt{F} \ket{1}_S + \sqrt{1-F} \ket{2}_S ) = e^{-i \phi} (\sqrt{F} \ket{\tilde 1}_{SM}+ \sqrt{1-F} \ket{\tilde 2}_{SM} ) = e^{i (\phi'-\phi)} \ket{\Psi_\sigma}_{SM}
\end{align}
Hence our machine is capable of preparing the two Gibbs state for 
\begin{align}
    F(\rho,\sigma) &= |\braket{\Psi_\rho}{\Psi_\sigma}|^2 =  |\braket{\psi_1(\tau)}{\psi_2(\tau)}|^2=\cos^2\left(\frac{\|\kappa\|  \tau }{2 \hbar}\right),\, \text{i.e.} \\
    \tau &= \frac{2 \hbar }{\|\kappa\|} \arccos |\braket{\Psi_\rho}{\Psi_\sigma}| = \frac{2 \hbar \, D(\sigma,\rho)}{\|H_S^{(2)}-H_S^{(1)}\|} =\frac{2 \hbar \, D(\omega(\beta, H_S^{(1)}),\omega(\beta, H_S^{(2)}))}{\|H_S^{(2)}-H_S^{(1)}\|} .
\end{align}
Of course it can also work for any longer time $\tau'>\tau$ by storing the two target states in a quantum memory after time $\tau$.

\section{Resonant level model}
\label{app:rlm}
Let us consider a single electron quantum dot with Hamiltonian
\begin{equation}
    H_S = E_S\hat a^\dagger \hat a~,
\end{equation}
where $\hat a^\dagger$ is its creation operator, it satisfies the anticommutation relation $\{\hat a^\dagger , \hat a\} = \id$ and $\{\hat a , \hat a\} = 0$. The energy gap of the two-level system is given by $E_S$. \\
Here we will model the ``thermalizing machine'' as a fermionic bath with Hamiltonian
\begin{equation}
    H_M =  \sum_{k=1}^n \Omega_k \hat b_k^\dagger \hat b_k~,
\end{equation}
$\hat b_k^\dagger$ is the creation operator of a bath mode with frequency $\Omega_k$. These satisfy the relations $\{\hat b_j^\dagger , \hat b_k\} = \delta_{jk}\id$, $\{\hat b_j , \hat b_k\} = \{\hat a , \hat b_k\} = \{\hat a^\dagger , \hat b_k\} = \{\hat a , \hat a\} = 0$. The machine interacts with the system via tunneling effects $\hat a^\dagger \hat b_k$ and $\hat b_k^\dagger \hat a$ with some modes of the machine, consistently with the most general quadratic Hamiltonian -- such as standard models describing quantum dots interacting with metallic leads.
All in all, the $S+M$ Hamiltonian writes as follows
\begin{equation}
    H_{tot}(t) = H_S + H_M + g(t) \sum_{k=1}^n \lambda_k \hat a^\dagger \hat b_k + \lambda_k^* \hat b_k^\dagger \hat a~. 
\end{equation}
Following the same steps as in Ref. \cite{Rolandi2023finitetimelandauer} (taking the wide-band and thermodynamic limit $\sum_{k}|\lambda_k|^2\delta(\Omega-\Omega_k)\rightarrow 2\pi$), one can solve the exact dynamics of the total system and compute the occupation probability of the quantum dot $p(t):= \Tr{\hat a^\dagger \hat a\rho(t)}$, we obtain
\begin{equation}\label{eqapp:prob_rlm}
    p(t) = p(0)e^{-\int_0^t\!dr g(r)^2} + \frac{1}{2\pi}\int_{-\infty}^{\infty}\!\!d\Omega~ f_\beta(\Omega)\left|\int_0^t ds~ g(s) e^{-\frac{1}{2}\int_s^t\!dr g(r)^2} e^{-i (E_S-\Omega)(t-s)}\right|^2~,
\end{equation}
where $f_\beta(\Omega) = (1+e^{\beta\Omega})^{-1}$ is the Fermi-Dirac distribution. For this system the thermal state is diagonal, and the norm of the off-diagonal element of $\rho(t)$ evolves as $e^{-\int_0^t\!dr g(r)^2}$. Since it decreases as fast or faster than $p(t)$ we will only consider initial states of the system which are diagonal.\\
\begin{figure}
    \centering
    \includegraphics[width=\linewidth]{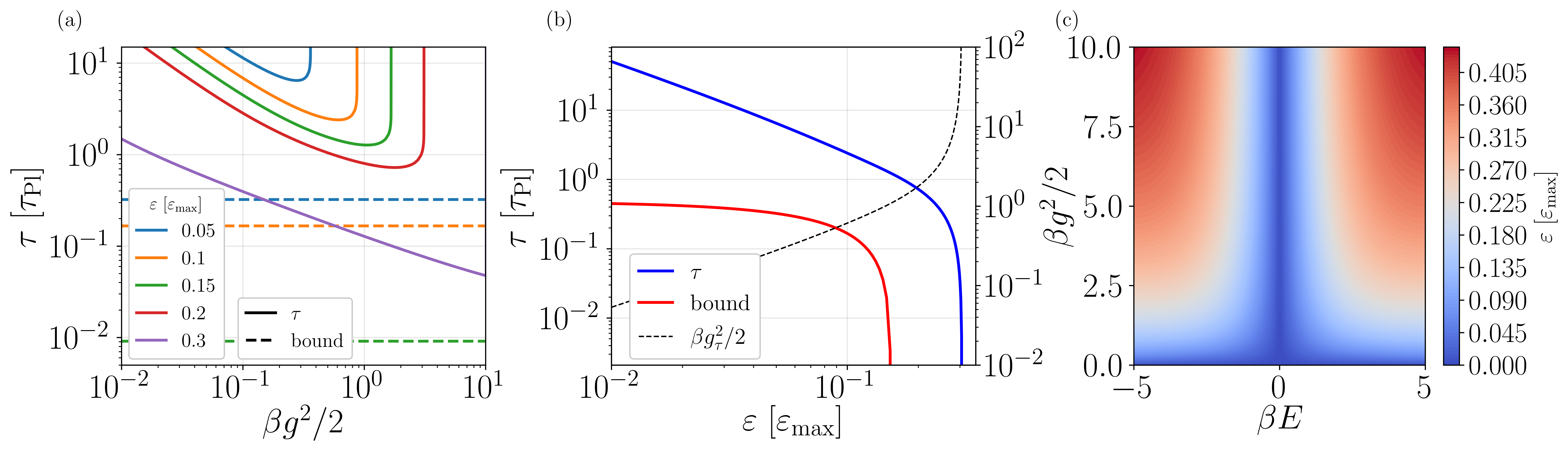}
    \caption{{\bf Thermalization machine with constant coupling.} Since for $p(0)=\frac{1}{2}$ and $E_S= 0$ this machine can instantly thermalize the system, Eq.~\eqref{eqapp:bound_2ham} imposes a bound on the thermalization speed for $E_S\neq0$. Here we fix $E_S^{(1)} = 0$ and $E_S^{(2)} = k_BT$ to illustrate the bound in different scenarios. By construction, for these energies, bound \eqref{eqapp:bound_2ham} becomes trivial for $\varepsilon \geq 0.153\varepsilon_{\mathrm{max}}$. {\bf (a)} Minimal thermalization time $\tau$ (plain lines) of the machine as a function of the coupling $g$ for multiple values of the accepted precision for thermalization. The dashed lines correspond to bound \eqref{eqapp:bound_2ham} of the corresponding $\varepsilon$, for $\varepsilon = 0.2\varepsilon_{\mathrm{max}}$ and $\varepsilon = 0.3\varepsilon_{\mathrm{max}}$ the bound is trivial (negative). Indeed we can see that the thermalizing machine is respecting the bound and that by reducing the precision of the machine we can achieve faster thermalization. At a given value of $g$ and above the machine fails to thermalize (the required time blow up to infinity) as the global thermal state it is converging to is too far from the local thermal state. {\bf (b)} Minimal thermalization time (blue) the machine can achieve by picking the optimal coupling as a function of the precision $\varepsilon$ (minima of the curves in (a)). The corresponding coupling $g_\tau$ that achieves the thermalization time $\tau$ is depicted with the dashed black line,on the other $y$-axis. The bound (Eq. \eqref{eqapp:bound_2ham}) is depicted by the red line. We can see that at $\varepsilon \approx 0.153\varepsilon_{\mathrm{max}}$ the bound becomes trivial, and that at $\varepsilon \approx 0.306\varepsilon_{\mathrm{max}}$ the thermalization time of the machine drops to zero because the initial state is within the machine precision. As we approach this limit, the optimal coupling diverges since the output state of machines with larger and larger couplings falls within the desired precision of thermalization ($\lim_{g\rightarrow\infty} \lim_{t\rightarrow\infty} p(t) = \frac{1}{2}$). 
    {\bf (c)} The precision required (i.e. the smallest value of $\varepsilon$) such that $\lim_{t\rightarrow\infty}p(t)$ (Eq. \eqref{eqapp:global_thermal}) is considered to be thermal, as a function of $E$ and $g$.
    }
    \label{fig:constant_g_machine}
\end{figure}

The main characterization we will investigate for this machine is the coupling $g(t)$. 
At time $\tau$ we disconnect the system from the machine and therefore we have $g(t)=0$ for $t\geq \tau$ and output $\rho(\tau)$. 
We will test how close we can get to the Planckian time bound for the set $\mathcal H = \{H^{(1)}_S, H^{(2)}_S\}$. In this case the Plackian bound Eq. \eqref{eq:chi_def} looses the max, we therefore have
\begin{equation}\label{eqapp:bound_2ham}
    \tau\geq \tau_{\mathrm{Pl}}\frac{ 2D(\omega_\beta(H^{(1)}_S), \omega_\beta(H^{(2)}_S)) -4\varepsilon}{\beta\|H_S^{(1)}-H_S^{(2)}\|}~.
\end{equation}

Before considering specific functions of time for $g(t)$, let us take the simpler case where it is a constant. In this case, we can solve the integrals in Eq.~\eqref{eqapp:prob_rlm} and show that $p(t)$ converges towards the thermal probability of the \emph{total} system as fast as $e^{-\frac{g^2}{2}t}$
\begin{equation}\label{eqapp:global_thermal}
    \lim_{t\rightarrow\infty} p(t) = \int_{-\infty}^{+\infty}\!\frac{d\Omega}{\pi}f_\beta(\Omega)\frac{g^2/2}{g^4/4+(E-\Omega)^2} = \frac{1}{2}-\frac{1}{\pi}\Im\Psi\!\left[\frac{1}{2}+\frac{\beta}{2\pi}\!\left(\!\frac{g^2}{2}+iE\!\right)\!\right]~,
\end{equation}
where $\Psi$ is the digamma function. This does not correspond to the desired thermal probability $f_\beta(E)$ (cf. Fig.~\ref{fig:constant_g_machine}c). Therefore, despite the fact that it might seem that we can thermalize the state as fast as we wish by increasing $g$, the machine will output a state that is further away from the desired one the faster it is. Since we allow for an error on the thermalizing machine, we can test the bound given by Eq.~\eqref{eqapp:bound_2ham} for this protocol. By picking $E_S^{(1)} = 0$ and $p(0) = \frac{1}{2}$ we find that the system's total (i.e. $S+M$) and local (i.e. only $S$) thermal state coincide and that the system does not evolve, therefore $D(\rho_S^{(1)}(t),\omega_\beta(H^{(1)}_S)) = 0$ for all $t\geq0$, which implies that the bound of Eq.~\eqref{eqapp:bound_2ham} is always true for $H^{(2)}_S$. We illustrate how this limits the performance of the machine to thermalize $H^{(2)}_S = k_BT\hat a^\dagger\hat a$ in Fig.~\ref{fig:constant_g_machine}.

\begin{figure}
    \centering
    \includegraphics[width=\linewidth]{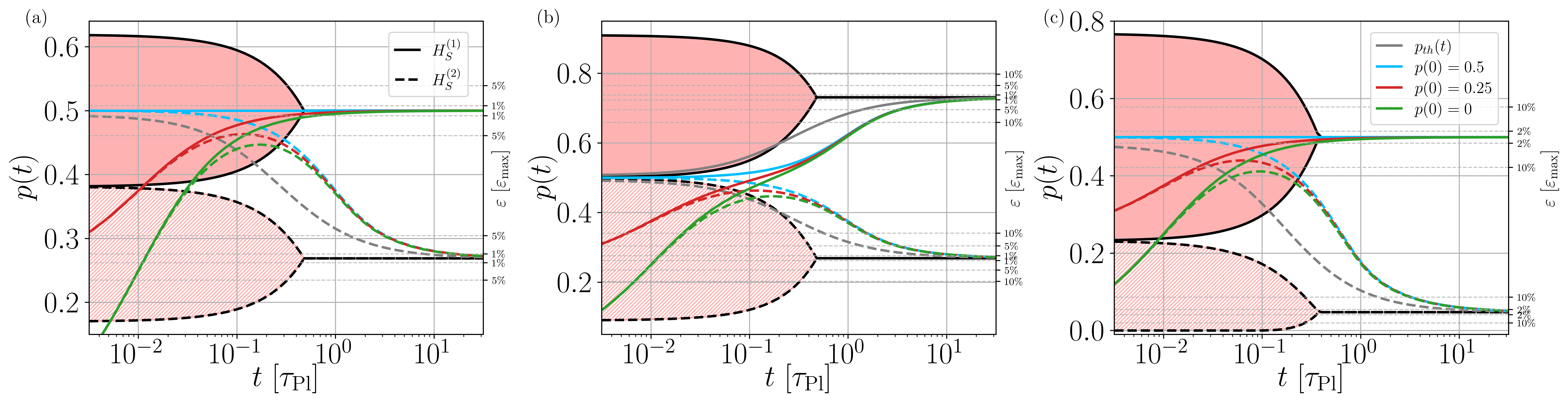}
    \caption{{\bf Thermalization machine with decaying coupling.}
    For two given Hamiltonians and the same initial condition, Eq. \eqref{eqapp:bound_turned} defines a forbidden area of the state space $(p^{(1)}(t), p^{(2)}(t))$ as a function of time. Here it is illustrated with the red-shaded areas: if a given machine thermalizes quickly one Hamiltonian $H_S^{(1)}$ (the plain line enters the plain red area at time $t$) then it cannot thermalize a second Hamiltonian $H_S^{(2)}$ (the dashed line cannot enter the dashed red area at time $t$) with arbitrary precision in the same amount of time. This must hold for any conceivable $M$ Hamiltonian that is able to thermalize $S$. Here, we illustrate for a machine with decaying coupling and $g(t) = \sqrt{a/(t+b)}$ for $a=\hbar$ and $b=\tau_{\mathrm{Pl}}/100$ and different choices of Hamiltonians: (a) $E_S^{(1)}=0$, $E_S^{(2)}=k_BT$ (b) $E_S^{(1)}=-k_BT$, $E_S^{(2)}=k_BT$  (c) $E_S^{(1)}=0$, $E_S^{(2)}=3k_BT$. The plain lines correspond to the evolution with $H_S^{(1)}$ and the dashed lines correspond to the evolution with $H_S^{(2)}$. The second $y$-axis marks the distance to the thermal state in units of the maximal Bures angle $\varepsilon_{\mathrm{max}}$.
    The colors represents different initial conditions, while the gray lines represent the instantaneous thermal probability $p_{th}(t)$ (Eq. \eqref{eqapp:global_thermal} with $g$ replaced by $g(t)$). The instantaneous thermal state does not correspond to any evolution, as the actual state will always lag behind it and therefore sets a bound on the evolution of the system. We have a tradeoff here where in the limit of $a$ small then the instantaneous thermal probability can violate the bound (as in (b), where $a$ is small for the specific configuration of the other parameters), but the coupling is so weak that the actual probability of the system is far from it because it is thermalizing slowly since the coupling is weak. While when $a$ is large we have that the instantaneous thermal state does not violate the bound (as in (a) and (c), where $a$ is large for the specific configuration of the other parameters), and therefore the state won't be able to do so either.
    }
    \label{fig:decaying_g_machine}
\end{figure}

Since the bound of Eq.~\eqref{eqapp:bound_2ham} is for the machine to thermalize \emph{all} of $\mathcal H$ within an error $\varepsilon$. We can turn the statement around into a speed-limit of evolution by taking $\varepsilon = D(\rho_S(t),\omega_\beta(H_S))$ and considering times smaller than the bound. There must be at least one of the Hamiltonians which has not managed to thermalize. Therefore we can re-arrange the bound to give a limit on the Bures distance between the state at time $t$, $\rho_S^{(i)}(\tau)$ (that evolves with $H^{(i)}_S + H_M + V_{SM}(t)$), and its corresponding thermal state $\omega_\beta(H^{(i)}_S)$
\begin{align}\label{eqapp:bound_turned}
    D(\rho_S^{(i)}(t),\omega_\beta(H^{(i)}_S))
    \geq \frac{1}{2}D(\omega_\beta( H^{(1)}_S),\omega_\beta(H^{(2)}_S)) - \frac{t}{4\tau_{\mathrm{Pl}}}\beta\|H_S^{(1)}-H_S^{(2)}\|
\end{align}
for at least one among $i=1$ and $i=2$. It is interesting to note that in this form of the bound the RSH only contains terms that depend on the system, while all the dependence on the thermalizing machine is on the LSH. We can see this bound in action in Fig.~\ref{fig:constant_g_machine} for a machine with constant coupling, where in panels (a) and (b) we showcase the time necessary to reach $\varepsilon = D(\rho_S(t),\omega_\beta(H_S))$. These results show very clearly that by increasing the coupling $g$ we do reduce the thermalization time-scale $2g^{-2}$, however as we do so the system thermalizes further and further away from the desired local thermal state. Therefore we can be as fast as we desire, however this comes at the cost of precision. \\

A way to combat the fact that the system thermalizes to the ``wrong'' state, is to change the coupling in time. Indeed, from Eq.~\eqref{eqapp:global_thermal} we can obtain $\lim_{g\rightarrow0}\lim_{t\rightarrow\infty} p(t) = f_\beta (E)$, which is the desired thermal output. Therefore we want to consider machines where the coupling is a function of time such that $g(t)$ drops to zero smoothly. However, let us note that the first term of Eq.~\eqref{eqapp:prob_rlm} is $p(0)e^{-\int_0^t\!dr g(r)^2}$, therefore if we pick a function $g(t)$ such that $\int_0^\infty\!dr g(r)^2 < \infty$ then we will have $e^{-\int_0^\infty\!dr g(r)^2} > 0$ implying that the system has not thermalized since the probability at $t=\infty$ is dependent on the initial probability $p(0)$. We can therefore conclude that for the system to thermalize exactly at $t=\infty$ the function $g(t)$ cannot decay faster than $t^{-1/2}$. Since we want the thermalization to happen as fast as possible, we study machines with functions of the form $g(t) = \sqrt{a/(t+b)}$. In this case we can simplify Eq.~\eqref{eqapp:prob_rlm} to 
\begin{equation}
    p(t) = \frac{1}{2} + \left(p(0)-\frac{1}{2}\right)\left(\frac{b}{t+b}\right)^a - \frac{1}{2\beta}\frac{a}{(t+b)^a}\int_0^t\!dsds'~ (s+b)^{\frac{a-1}{2}}(s'+b)^{\frac{a-1}{2}}\frac{\sin(E(s-s'))}{\sinh(\pi(s-s')/\beta)}~.
\end{equation}
From this we can compare the performance of this thermalizing machine to the bound given by Eq.~\eqref{eqapp:bound_turned}, we look at a few examples and discuss them in Fig.~\ref{fig:decaying_g_machine}.

\end{document}